\documentclass[aps,showpacs,showkeys,nofootinbib,twocolumn]{revtex4}
\usepackage{epsfig,amssymb,bm,graphics,color}
\newcommand*{\fs}[1]{#1\!\!\!/}

\newcommand*{\ee}{e^+e^-}

\begin{document} {\normalsize }

\title{Non-linear Breit-Wheeler process in %two consecutive
short laser double-pulses}
%
%%%%%%%%%%%%%%%%%%%%%%%%%%%%%%%%%%%%%%%%%%%%%%%%%%%
 %%%
 \author{A.~I.~Titov}
 \affiliation{
 Bogoliubov Laboratory of Theoretical Physics, JINR, Dubna 141980, Russia}
\author{H.~Takabe}
 \affiliation{
 Helmholtz-Zentrum  Dresden-Rossendorf, 01314 Dresden, Germany}

\author{B.~K\"ampfer}
 \affiliation{
 Helmholtz-Zentrum  Dresden-Rossendorf, 01314 Dresden, Germany}
 \affiliation{
 Institut f\"ur Theoretische Physik, TU~Dresden, 01062 Dresden, Germany}

\begin{abstract}
 The non-linear (strong-field) Breit-Wheeler $\ee$ pair production by
a probe photon traversing
  two consecutive short and ultra short (sub-cycle) laser pulses
  is considered within a QED framework.
  The temporal shape of the pulses and the distance between
  them are essential for the differential cross section as a function
  of the azimuthal angle distribution of the outgoing electron (positron).
  The found effect of a pronounced azimuthal anisotropy
is important for sub-cycle pulses and decreases rapidly
  with increasing width of the individual pulses.
\end{abstract}


 \maketitle

\section{Introduction}

 The rapidly progressing laser technology \cite{Tajima}
 offers novel and unprecedented opportunities to investigate
 quantum systems with intense laser beams~\cite{Piazza}.
 An intensity $I_L$ of $\sim 2\times 10^{22}$  W/cm${}^2$ has been already
 achieved~\cite{I-22}. Intensities of the order of
 $I_L \sim 10^{23}...10^{25}$ W/cm$^2$ are envisaged in the near future, e.g.\
 at  CLF~\cite{CLF}, ELI~\cite{ELI}, or HiPER~\cite{hiper}.
 Further facilities are in the planning or construction stages, e.g.
 the PEARL laser facility~\cite{sarov} at Sarov/Nizhny Novgorod, Russia.
 The high intensities are provided in short
 pulses on a femtosecond pulse duration
 level~\cite{Piazza,ShortPulse,ShortPulse_2},
 with only a few oscillations of the electromagnetic (e.m.) field
 or even sub-cycle pulses.
 (The tight connection of high intensity and short pulse duration
 is further emphasized in \cite{Mackenroth-2011}. The attosecond
 regime will become accessible at shorter wavelengths~\cite{atto,I-222}).

 New laser facilities may utilize
 short and ultra-short pulses in "one-" or "few-" cycle regimes.
 In this case, a determination of the pulse fine-structure is very important
 and, in particular, tasking the phase difference between the electric
 field and pulse envelope, i.~e. the carrier envelope phase (CEP).
 It was found that the CEP effect is especially important just for the case of the
 short and ultra-short (sub-cycle) pulses
 (cf.~\cite{CEPTitov} and references therein;
for recent access option for the CEP in long pulses, cf.~\cite{Li:2018xnp}).

 The study of quantum processes in two consecutive
(or double) laser pulses
 with taking into account CEP effect is a new important and
 interesting topic in laser physics.
 An analysis of Breit-Wheeler pair production within the
framework of scalar electrodynamics is provided in \cite{JansenMuller}.
 Below, we give a further development of this problem.
 We extend that previous consideration to QED, where
 $\ee$ are fermions, and we concentrate our attention
 on the azimuthal angle distribution of outgoing electron (positron)
 at fixed CEP. In fact, the non-linear Breit-Wheeler process considered below,
 assumes the interaction of a probe photon $X$
(e.g.\ from Compton backscattering in a pre-pulse, starting
a seeded cascade via two-step part of trident process,
cf.~\cite{Blackburn:2017dpn,Blackburn:2018ghi})
with two consecutive laser pulses, $L_{1+2}$, in the reaction
 $X+L_{1+2} \to e^+ +e^-$, where a multitude of laser photons
 can participate simultaneously in the $\ee$ pair creation,
thus enabling the process even "below threshold".
In the Furry picture, this process
is a cross channel of the non-linear Compton process without having
a classical counterpart, as the Thomson process in the weal-field regime.
The non-linear Breit-Wheeler process means the decay of the probe
photon into a laser-dressed electron ($e^-$) positron ($e^+$) pair.
 The emphasis here is on short and intense laser pulses.
 Long and weak laser pulses are dealt with in the standard
 textbook Breit-Wheeler process (cf.\ the review paper \cite{Ritus-79}).
 Various aspects, such as the impact of
the pulse envelope, % \cite{...},
pulse duration, % \cite{...},
pulse polarization, % \cite{...},
of $\ee$ pair creation in single pulses were analyzed, e.g.\
 in Refs.~\cite{TitovPEPAN,TitovPRA,TitovPRL,Nousch,Krajewska}
with special emphasis on
CEP effects \cite{A1,CEPTitov},
spin effects \cite{Jansen:2016gvt},
bi-frequent pulses \cite{Jansen:2013dea,Nousch:2015pja,Otto:2016fdo},
spectral caustics \cite{Nousch:2015pja},  and
focusing effects \cite{DiPiazza:2016maj,DiPiazza:2016tdf}.

One may contrast the single-pulse laser beams to a long train of pulses or
multi-pulses, where a special modulation of the phase space distribution
of produced $e^\pm$ in comb structures arises due to interference effect \cite{Krajewska:2014ssa}.
In some sense, such a situation refers to multi-shot experiments, where
a laser with extremely high repetition rate fires for some time.
(Another option is the pulse train generation at XFELs,
cf.~\cite{Decker}.)
In between is the presently considered case of a double pulse.
Below, we concentrate mainly on the interplay of the effect
 of  CEP and the shape of the laser beams which is
 determined by the temporal shape of the individual pulses (cf.~\cite{TitovPRA})
 and the separation distance between them.

Our present paper is a follow-up of \cite{CEPTitov} which is inspired by
\cite{JansenMuller}. Differences are
(i) Fermion pairs (\cite{JansenMuller} deals with Bose pairs),
(ii) circular polarization (\cite{JansenMuller} uses linear polarization),
(iii) many details of the laser pulse modeling (\cite{JansenMuller} uses
separate individual pulses with particular envelope functions, while our field
arises from "cutting out" the pulses by a double hump window function
with hump separation $\ge 0$),
(iv) consideration of very short and sub-cycle pulses,
(v) focus on sub-threshold pair production,
(vi) focus on azimuthal angle distribution (\cite{JansenMuller} considers
the energy spectra).

 Our paper is organized as follows.
In Sect.~II, the double-pulse field model is presented.
 Sect.~III recalls the basic expressions  for the relevant observables
 in non-linear Breit-Wheeler $\ee$ pair production.
 In Sect.~IV we discuss results of
 numerical calculations. Our summary is given in Sect.~V.

\section{Model of the double pulse}

 In the following we use the electromagnetic (e.m.) four-potential
 for a circularly polarized laser field in
 the axial gauge $A^\mu=(0,\,\mathbf{A}(\phi))$ with
\begin{eqnarray}
 \mathbf{A}(\phi)=f(\phi) \left( \mathbf{a}_1\cos(\phi+\tilde\phi)+ \mathbf
 {a}_2\sin(\phi+\tilde\phi)\right)~,
  \label{III1}
 \end{eqnarray}
 where $\tilde\phi$ is the CEP. The quantity
 $\phi=k\cdot x$ is the invariant phase with four-wave vector
 $k=(\omega, \mathbf{k})$, obeying the null field property
$k^2 = k \cdot k=0$\footnote{Effects of an ambient medium,
e.g.\ a plasma, can be accommodated in a modified dispersion
relation, $k^2 \ne 0$ \cite{Mackenroth:2018rtp}, as customary
done in many phenomenological QCD approaches,
e.g.\ in \cite{Kampfer:1999ff}.}
 (a dot between four-vectors indicates the Lorentz scalar
 product) implying $\omega = \vert\mathbf{k}\vert$,
 $ \mathbf{a}_{(1,2)} \equiv \mathbf{a}_{(x,y)}$;
 $|\mathbf{a}_x|^2=|\mathbf{a}_y|^2 = a^2$, $\mathbf{a}_x \mathbf{a}_y=0$;
 transversality means $\mathbf{k} \mathbf{a}_{x,y}=0$ in the present gauge.
 The envelope function $f(\phi)$ in case of two consecutive pulses is
 chosen as sum of two hyperbolic secants
 \begin{eqnarray}
 f(\phi) %\equiv f_1(\phi)+f_2(\phi)
=\frac{1}{\cosh(\frac{\phi}{\Delta}-G)}
 + \frac{1}{\cosh(\frac{\phi}{\Delta}+G)}
 \label{III2},
 \end{eqnarray}
 where $G$ is the separation parameter of two consecutive short pulses
 which is equal to the distance between the centers of the two
 consecutive pulses.
 The dimensionless quantity $\Delta$ is related to the single pulse duration
 $2\Delta=2\pi N$, where $N$ has the meaning of a number of cycles in
 the individual pulse.
 It is related to the time duration of the pulse $\tau=2N/\omega$.

One may imagine the generation of the above described pulse structures
by the sequence of a laser beam - beam splitter -
a delay section for one of the split beams and
subsequently the merging of both beams.
Presently, the alignment of two separate laser beams of seems hardly possible
to be realized with equal carrier envelope phases and precisely adjustable
temporal pulse delay together with keeping constant the polarizations.
In this respect, our field model (1, 2) is highly idealized, but enjoys a minimum
number of parameters. However, having in mind principle effects of quantum
interference patterns by such double-slit phenomena in the temporal domain,
our field model can be regarded as useful representative ansatz. For further
discussion of aspects of double-and multi-slit phenomena we refer the interested
reader to \cite{JansenMuller} with references quoted therein.

\begin{figure}[tb]
 \includegraphics[width=0.48\columnwidth]{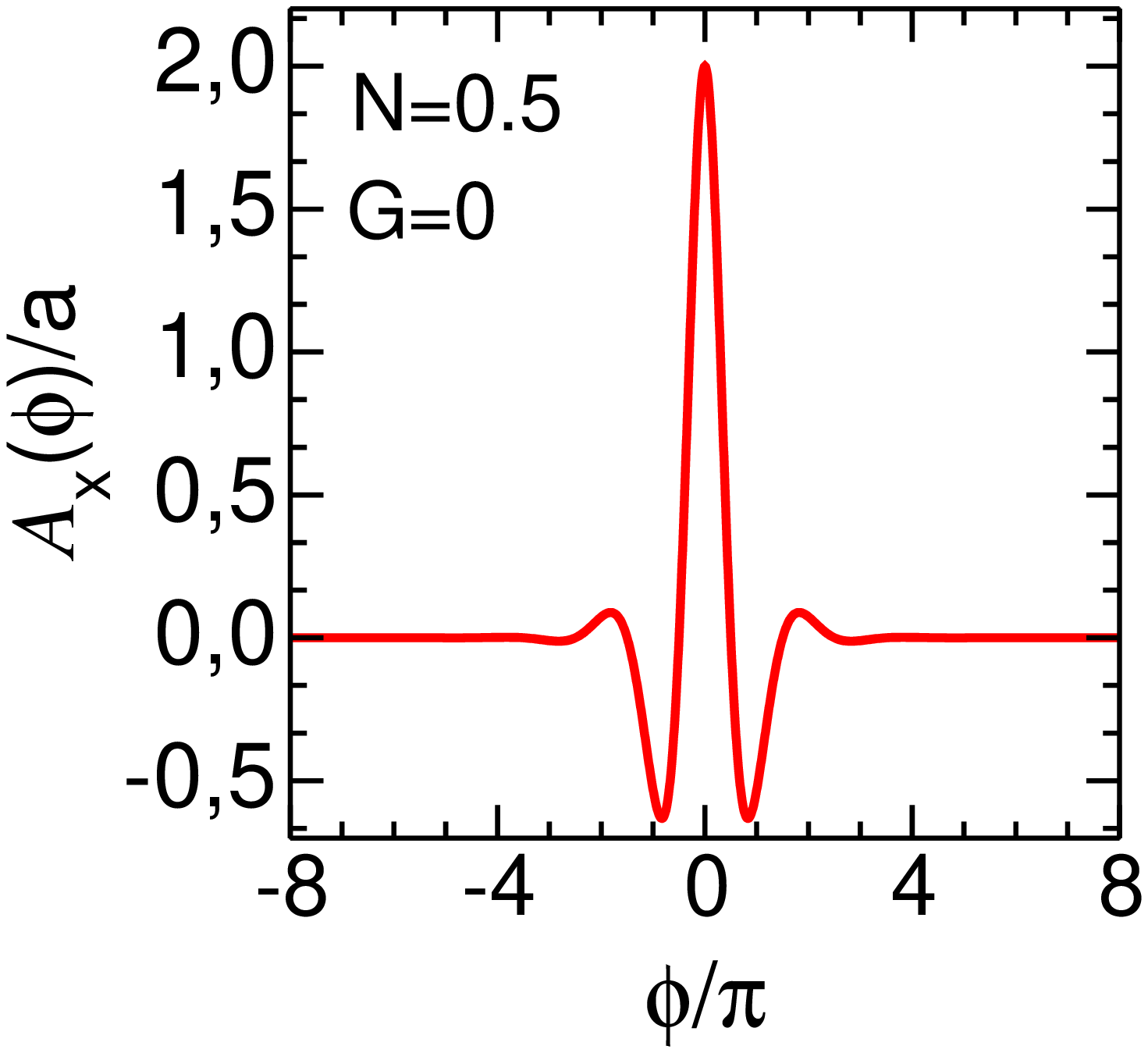} \hfill
 \includegraphics[width=0.48\columnwidth]{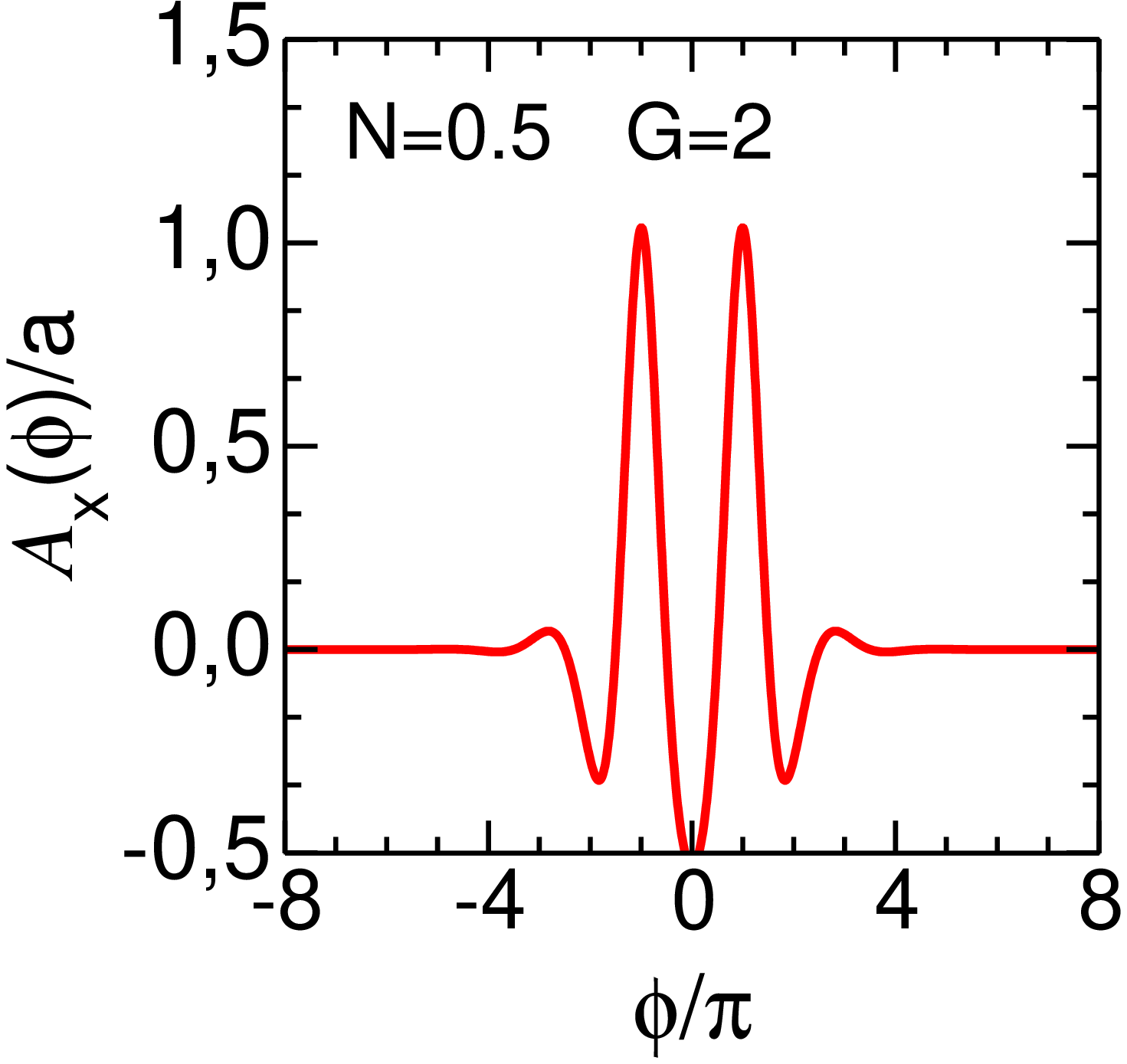} \\
 \includegraphics[width=0.48\columnwidth]{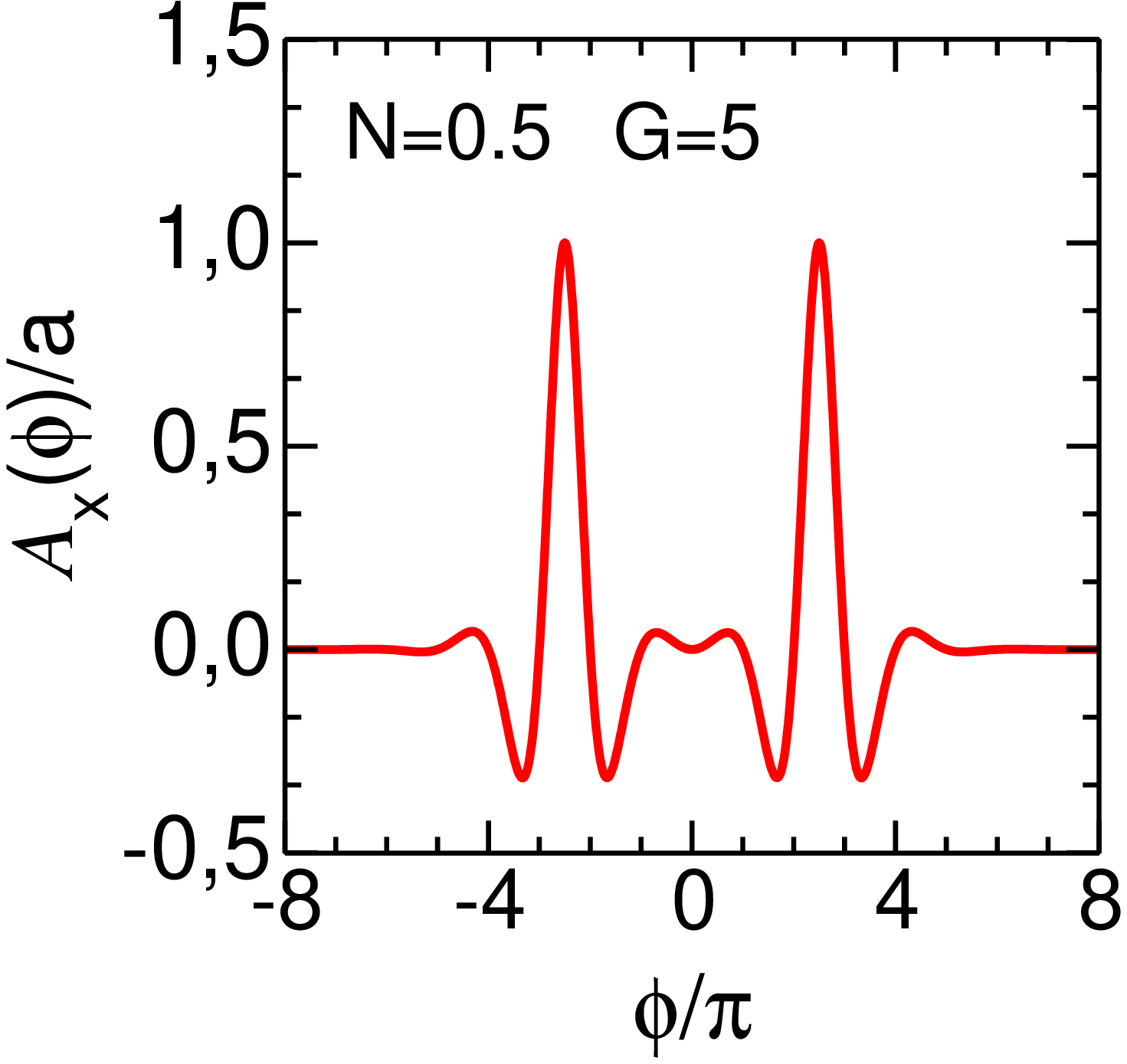} \hfill
 \includegraphics[width=0.48\columnwidth]{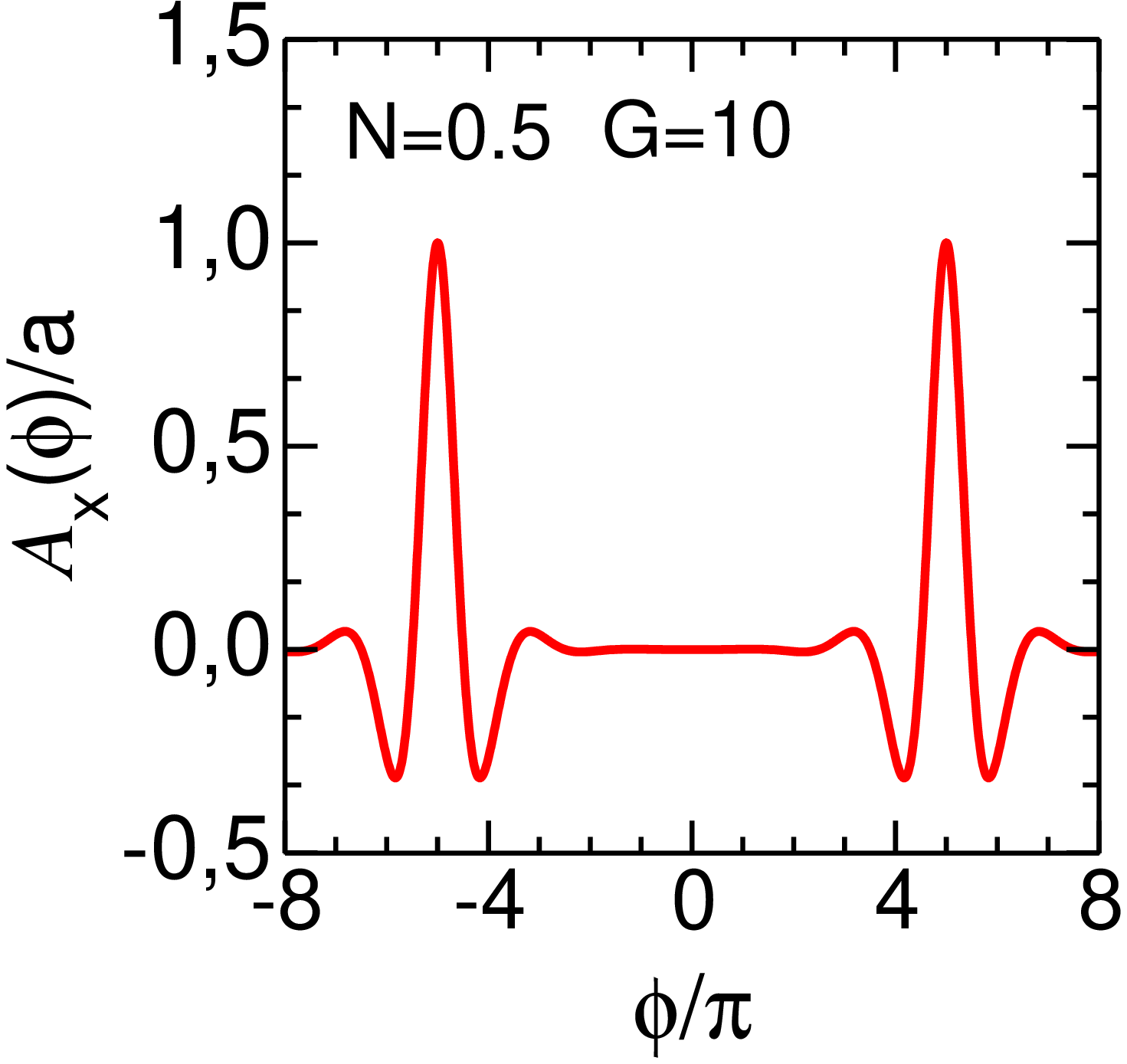}
 \caption{\small{(Color online)
 The e.m.\ potential $A_x/a$ in Eq.~(\ref{III2})
 as a function of the invariant phase $\phi=k \cdot x$
 for sub-cycle pulse with $N = 1/2$ and different values
 of the separation parameter  $G=0,\,2,\,5,\,10$ as
 it is indicated in the legends.
 \label{Fig:01}}}
 \end{figure}

 For an illustration, in Fig.~\ref{Fig:01} we show the $x$ component
 of the e.m.\ potential $A_x/a$ as a function of
 invariant phase $\phi$ for different values of the separation parameter $G$.
 The case $G=0$ corresponds to the complete overlap of the two envelopes
 which leads to the e.m.\ potential with a double amplitude $a$.

 \begin{figure}[tb]
 \includegraphics[width=0.48\columnwidth]{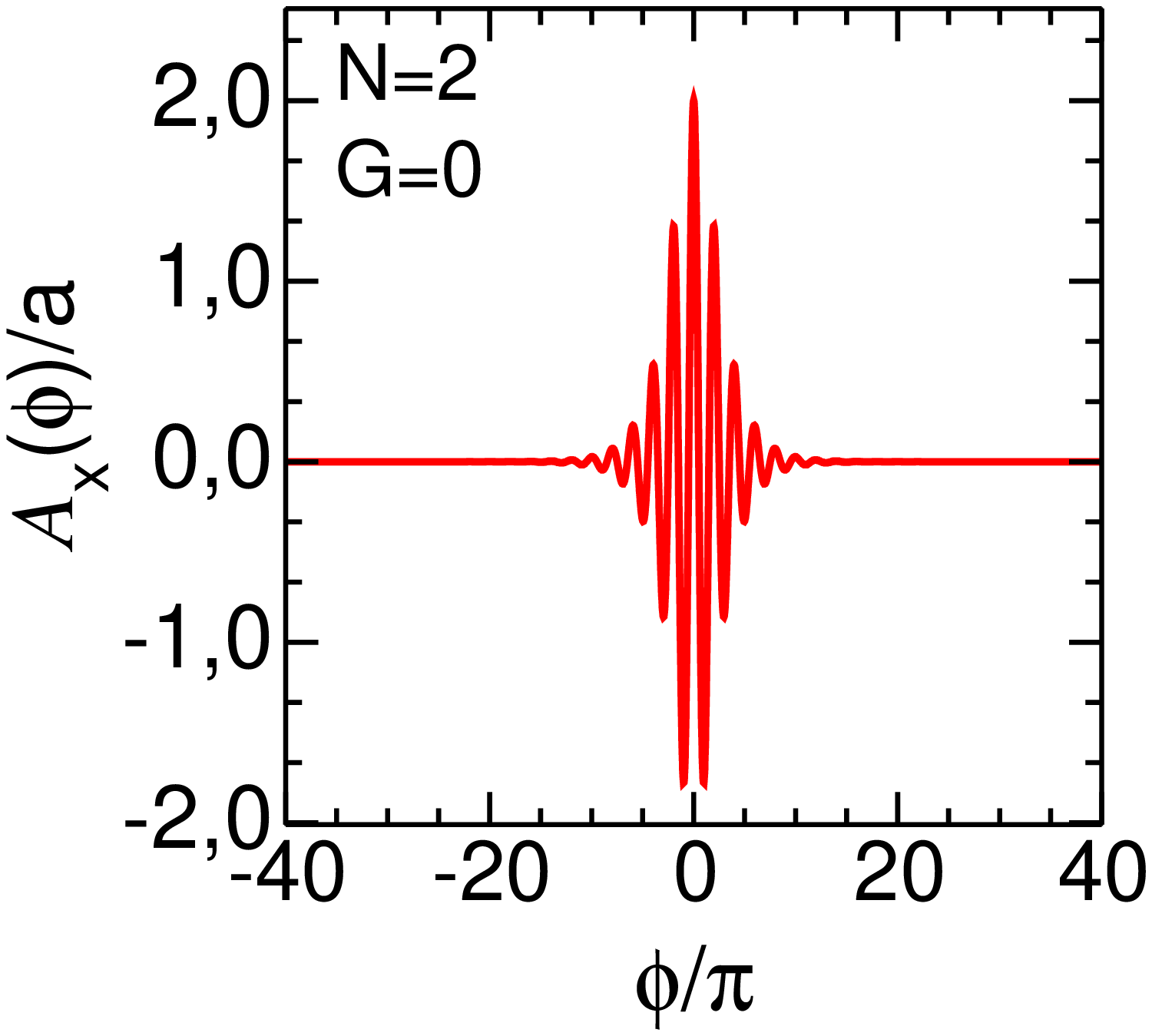} \hfill
 \includegraphics[width=0.48\columnwidth]{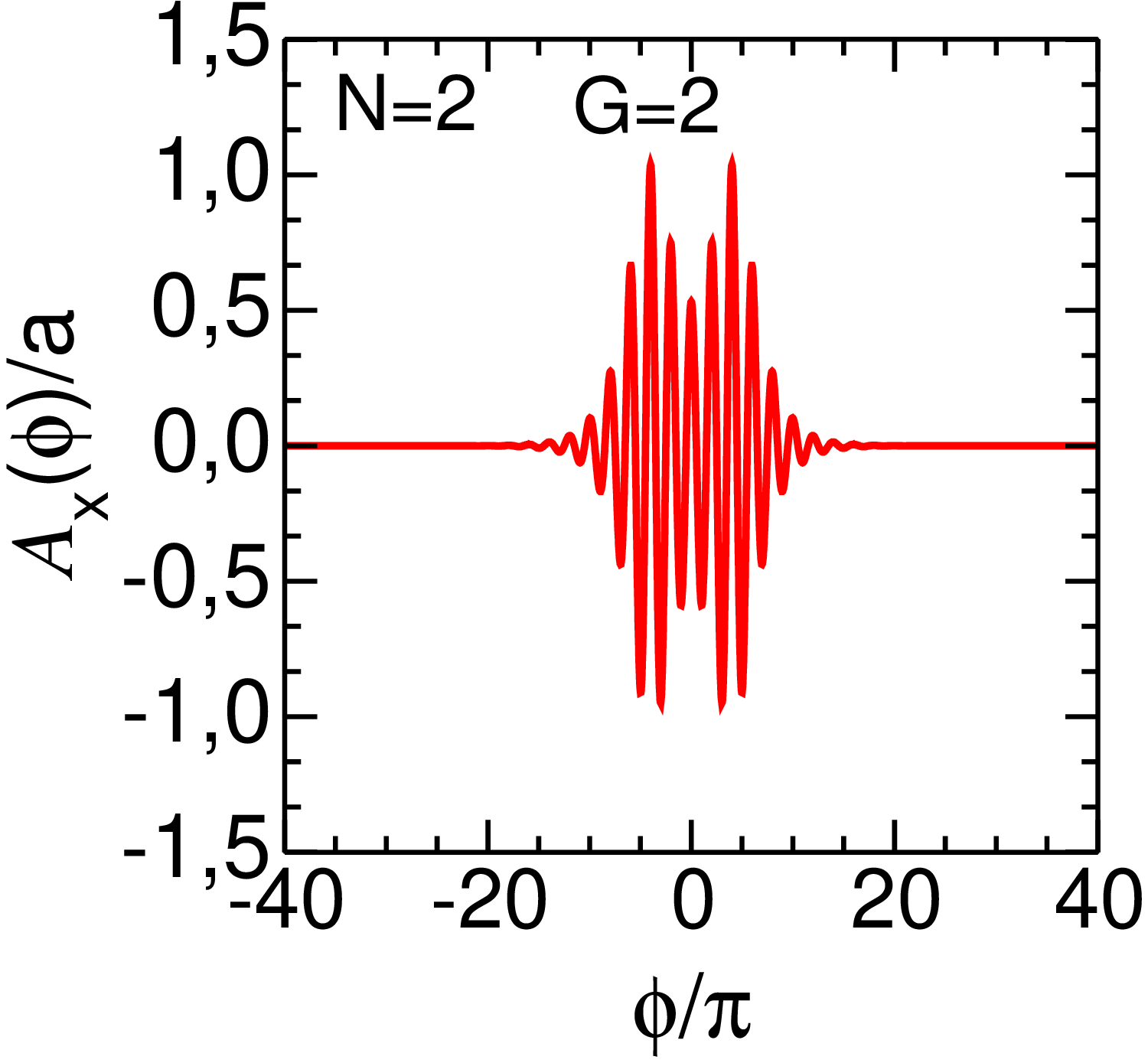} \\
 \includegraphics[width=0.48\columnwidth]{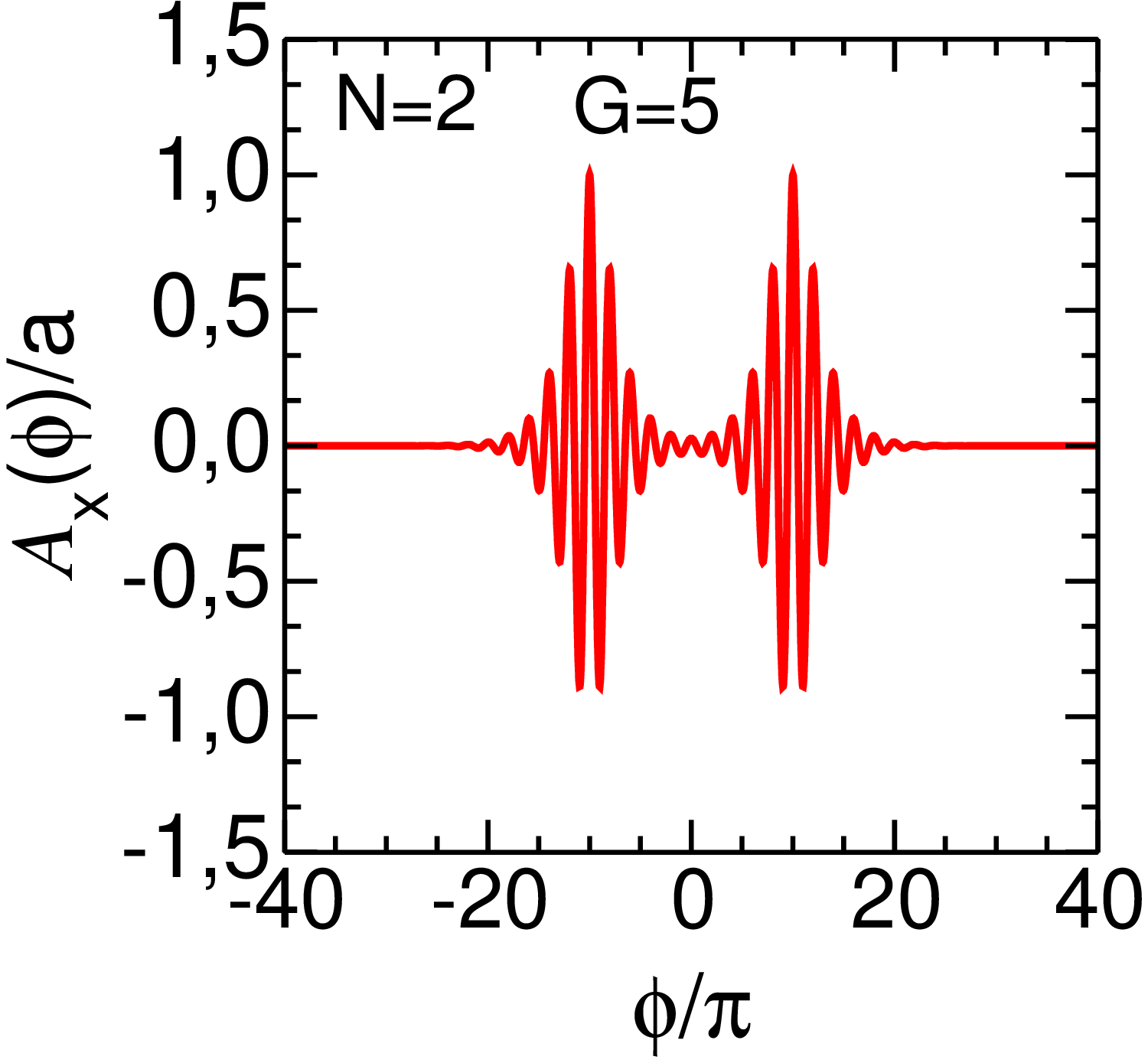} \hfill
 \includegraphics[width=0.48\columnwidth]{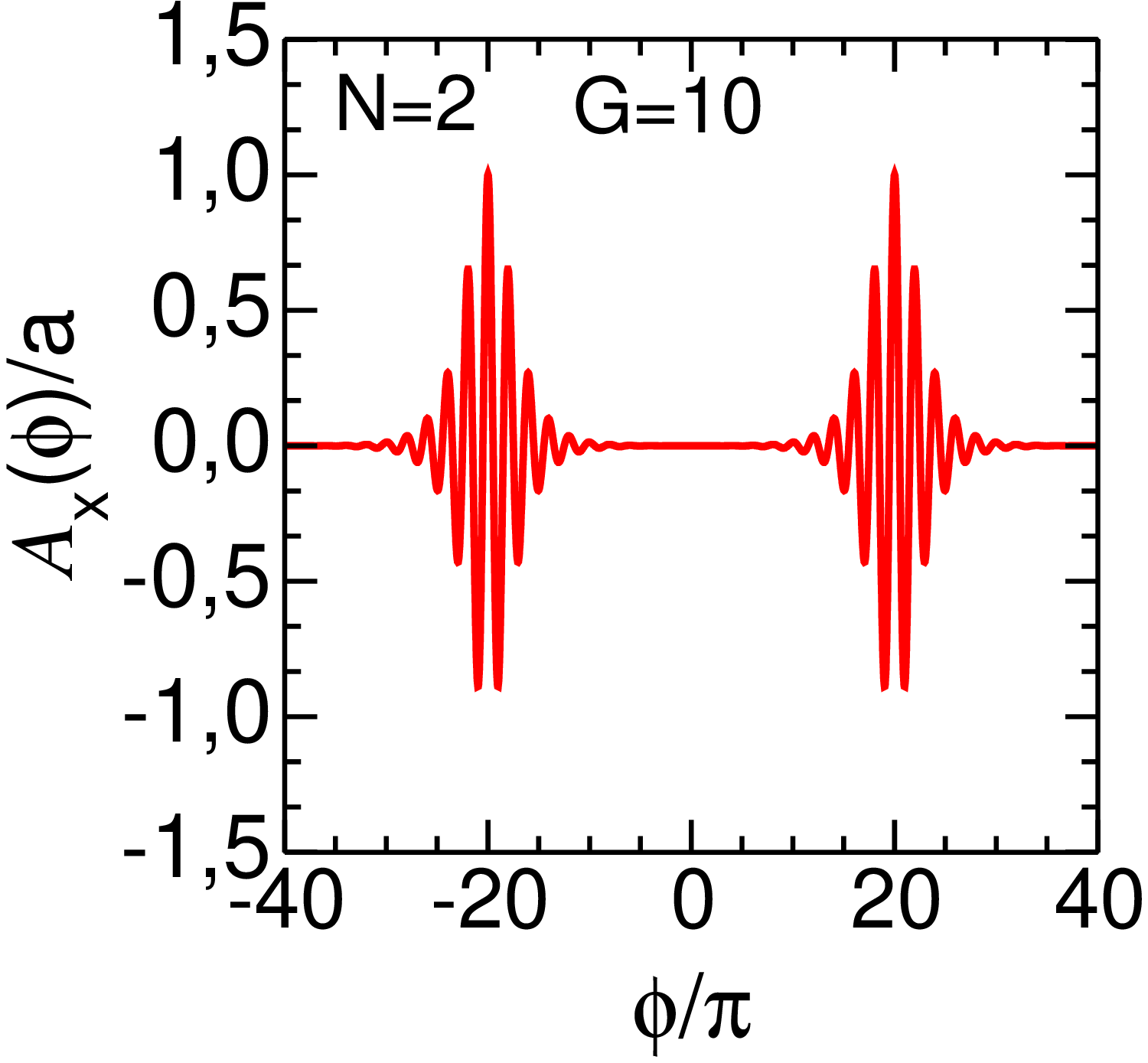}
 \caption{\small{(Color online)
 The same as in Fig.~\ref{Fig:01} but for a short pulse with $N=2$.
 \label{Fig:02}}}
 \end{figure}

The e.m.\ potential $A_x/a$ for a short pulse with $N=2$ and different $G$
 is exhibited in Fig.~\ref{Fig:02}.

 As mentioned above, the interplay between carrier
 envelope phase $\tilde\phi$ and  separation parameter $G$
 is the main subject of our present discussion
 and, as we will show, it has a strong impact on the azimuthal angle distribution
 of the outgoing electron (positron), in particular for a short pulse duration $\Delta$.
 We will consider essentially multi-photon events, where
 a finite number of laser photons is involved into
 the $\ee$ pair production. This allows for the sub-threshold
 $\ee$ pair production
 with $s<s_{\rm thr}$, where $s = (k_x + k)^2 $ is the square of total energy in
 the center of mass system (c.m.s., defined by $\mathbf k_x = - \mathbf k$,
$s_{c.m.s.} = 4 \omega_X \omega$),
$s_{\rm thr}=4m^2$ is it threshold value,
 where $m$ is the electron mass. We also discuss dependence of the
 cross section on e.m. field intensity which is described
 by the reduced field intensity
 $\xi^2=e^2a^2/m^2$. We use natural units with
 $c=\hbar=1$, $e^2/4\pi = \alpha \approx 1/137.036$.

\section{Cross section and anisotropy}

The azimuthal angular differential cross can be cast into the form
(cf.\ Appendix A for details)
\begin{eqnarray}
\frac{d \sigma}{d\phi_{e^+} }
=\frac{\alpha^2\,\zeta}{4m^2\xi^2 N_0}
\,\int\limits_\zeta^{\infty}\,d\ell \,v(\ell) \,
\int\limits_{-1}^{1} d\cos\theta_{e^+}
\, w{(\ell)}~
\label{III99}
\end{eqnarray}
with partial probabilities
\begin{eqnarray}
\label{III26-0}
 w(\ell) &=& 2 |\widetilde Y_\ell(z)|^2+\xi^2(2u-1)  \\
& \times  &\left(|Y_{\ell-1}(z)|^2 + |Y_{\ell+1}(z)|^2 -2
{\rm Re}\,(\widetilde Y_\ell(z)X^*_\ell(z))\right), \nonumber
\end{eqnarray}
which recover the known expressions \cite{Ritus-79} in case of
infinitely long e.m.\ pulses, where $\ell$ becomes a discrete (integer)
variable (cf.\ \cite{CEPTitov}).
The azimuthal angle of the
 outgoing positron, $\phi_{e^+}$, is defined as
 $\cos\phi_{e^+}={\mathbf a_x}{\mathbf p}_{e^+}/a |{\mathbf p}_{e^+}|$.
 It is related to the azimuthal angle of the electron by
 $\phi_{e}=\phi_{e^+} - \pi$.
Furthermore,  $\theta_{e^+}$ is the polar angle
 of outgoing positron, $v$ is the positron (electron) velocity
 in c.m.s.. The averaging and sum over
 the spin variables in the initial and the final states is executed.

 The lower limit of the integral over the variable $\ell$
 is the threshold parameter $\zeta=s_{\rm thr}/s\equiv4m^2/s$.
 The region of $\zeta<1$
 corresponds to the above-threshold $\ee$ pair production, while
 the region of $\zeta>1$ matches the sub-threshold pair production.
We keep our notation of \cite{CEPTitov} and
denote four-vectors $k(\omega,{\mathbf k})$,
 $k'(\omega',{\mathbf k}') \equiv k_x(\omega_x,{\mathbf k}_x)$,
 $p(E,{\mathbf p})$ and $p'(E',{\mathbf p}')$ as the four-momenta of the
 background (laser) field (\ref{III1}),
 incoming probe photon, outgoing positron and electron,
 respectively.  The variables $s$, $v$ and $u$
 are determined by $s={2k\cdot k' }= 2(\omega'\omega -{\mathbf k}'{\mathbf k})$
(with $\mathbf k' \mathbf k = - \omega' \omega$ for head-on geometry),
 $v^2=(\ell s-4m^2)/\ell s$,
 $u\equiv(k'\cdot k)^2/\left(4(k\cdot p)(k\cdot p')\right)=1/(1-v^2\cos^2\theta_e)$.
 The factor $N_0$ reads
$ N_0={1}/{2\pi}\int_{-\infty}^{\infty}
 d\phi\,(f^2(\phi)+ {f'}^2(\phi)) $
 and normalizes to the photon flux in case of finite pulses \cite{TitovEPJD}.
 The variable $\ell$ takes continuous values and the product
 $\ell\omega$ has the meaning of the laser energy
 involved in the process (see also~\cite{A1} for a recent discussion).

The basic functions $Y_\ell$ and $X_\ell$ entering partial probabilities (\ref{III26-0})
the may be considered as generalized Bessel functions for the finite e.m.\ pulse,
\begin{eqnarray}
Y_\ell(z)&=&\frac{1}{2\pi} {\rm e}^{-i\ell(\phi_0-\tilde\phi)}\int\limits_{-\infty}^{\infty}\,
d\phi\,{f}(\phi)
\,{\rm e}^{i\ell\phi-i{\cal P}(\phi)} ~, \label{Yl}\\
X_\ell(z)&=&\frac{1}{2\pi}{\rm e}^{-i\ell(\phi_0-\tilde\phi)} \int\limits_{-\infty}^{\infty}\,
d\phi\,{f^2}(\phi)
\,{\rm e}^{i{\ell} \phi-i{\cal P}(\phi)}, \label{Xl}\\
\widetilde Y_\ell(z)&=&\frac{z}{2\ell} \left(Y_{\ell+1}(z) +
Y_{\ell-1}(z)\right) - \xi^2\frac{u}{u_\ell}\,X_\ell(z)
\label{CP3}
\end{eqnarray}
 with the phase function
 \begin{eqnarray}
{\cal P(\phi)}&=&z\int\limits_{-\infty}^{\phi}\,d\phi'\,
\cos(\phi'-\phi_0+\tilde\phi)f(\phi')\nonumber\\
&-&\xi^2\zeta u\int\limits_{-\infty}^\phi\,d\phi'\,f^2(\phi')~,
\label{CP2}
\end{eqnarray}
where $\phi_0$ is the azimuthal angle of the outgoing electron
$\phi_0 \equiv \phi_e=\phi_{e+} - \pi$
and the {argument $z$ of the generalized Bessel functions} is related to
$\xi$, $\ell$ and $u$ via
$ z=2{\ell}\xi\sqrt{{u}/{u_\ell}\left(1-{u}/{u_\ell}\right)}$
with $u_\ell\equiv\ell/\zeta$.

 Together with the differential cross section $d\sigma / d\phi_e$
 we analyze the anisotropy of the electron (positron) emission
 as a function as a function of the positron (electron)
 azimuthal angle
  \begin{eqnarray}
{\cal A}=
 \frac{{d\sigma(\phi_{e})}
-{d\sigma(\phi_{e^+})}}
{{d\sigma(\phi_{e})}
+{d\sigma(\phi_{e^+})}} .
\label{U9}
\end{eqnarray}
%with $\phi_e=\phi_{e^+} - \pi$.

The cross section (\ref{III99}) and anisotropy (\ref{U9}) have
a non-trivial, non-monotonic dependence
as a function of azimuthal angle $\phi_{e^+}$ $(\phi_{e})$,
which is determined by the values of the
carrier phase $\tilde\phi$ and separation parameter $G$.
% in the envelope function~(\ref{III2}).
On a qualitative level, the reason for such behavior is the following.
The basic functions $Y_\ell$ and $X_\ell$ are determined by the integral
 over $d\phi$ with a rapidly oscillating exponential function $\exp[i\Psi]$ with
leading terms
\begin{eqnarray}
\Psi= \ell\phi
 &-& z\Delta \cos(\Phi-G\Delta)\int\limits_{-\infty}^{\frac{\phi}{\Delta}-G} dt\, f_0(t)\cos(t\Delta)
 \nonumber\\
&-& z\Delta  \sin(\Phi-G\Delta)\int\limits_{-\infty}^{\frac{\phi}{\Delta}-G} dt\, f_0(t)\sin(t\Delta)
\nonumber\\
 &-& z\Delta \cos(\Phi+G\Delta)\int\limits_{-\infty}^{\frac{\phi}{\Delta}+G} dt\, f_0(t)\cos(t\Delta)
\nonumber\\
 &-& z\Delta  \sin(\Phi+G\Delta)\int\limits_{-\infty}^{\frac{\phi}{\Delta}+ G} dt\, f_0(t)\sin(t\Delta)~, \nonumber\\
&+& \cdots
 \label{UU8}
 \end{eqnarray}
 where $f_0(t)=1/\cosh(t)$ and $\Phi=\phi_e-\tilde\phi$.
The ellipses refer to contributions from the second line in (\ref{CP2}).
 Due to the fact that for short (ultra-short) pulses
 \begin{eqnarray}
 \int\limits_{-\infty}^{t} dt'\, f_0(t')\cos(t'\Delta)
 \gg
 \int\limits_{-\infty}^{t} dt'\, f_0(t')\sin(t'\Delta)~,
 \label{UU81}
\end{eqnarray}
 one can conclude that the maximum value to the highly oscillating integrals
 determined the basic functions $Y_\ell$ and $X_\ell$ comes from the range
 $\Phi \pm G\Delta\simeq 2 n \pi$, where $n$ is an integer $n=0, \,1, \,2\, \cdots$.
 Then the differential cross section would be enhanced at
 \begin{eqnarray}
 \phi^{\rm max}_{e^{+}}-\tilde\phi=(2n+1)\pi\pm G\Delta~.
 \label{phi1}
 \end{eqnarray}

\section{Numerical results}

\subsection{Azimuthal distributions}

Let us consider first the sub-cycle pulse with $N=1/2$, i.e.\ $\Delta=\pi/2$.
For sub-cycle pulses, the electric field field vector does not rotate
in the $\mathbf k$-transverse plane with full length. This asymmetry,
which is also seen in Fig.~1 as strong asymmetry in the $\pm A_x$
directions, leaves such a strong imprint on the azimuthal distribution
of $e^\pm$. It rapidly disappears for longer pulses, i.e. larger values
of $N$, as already evidenced by Fig.~2.

\begin{figure}[htb]
\includegraphics[width=0.48\columnwidth]{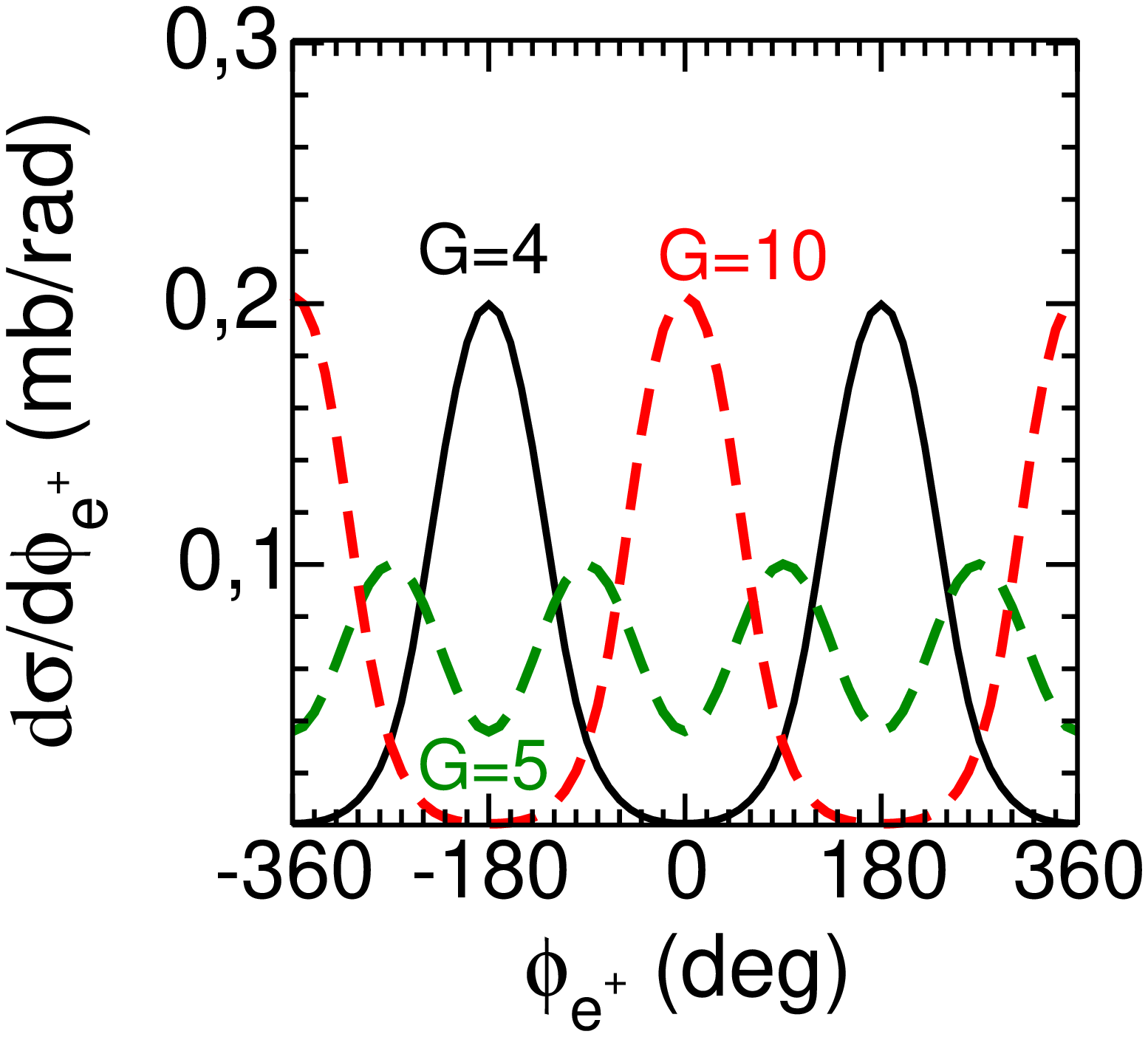}  \hfill
\includegraphics[width=0.48\columnwidth]{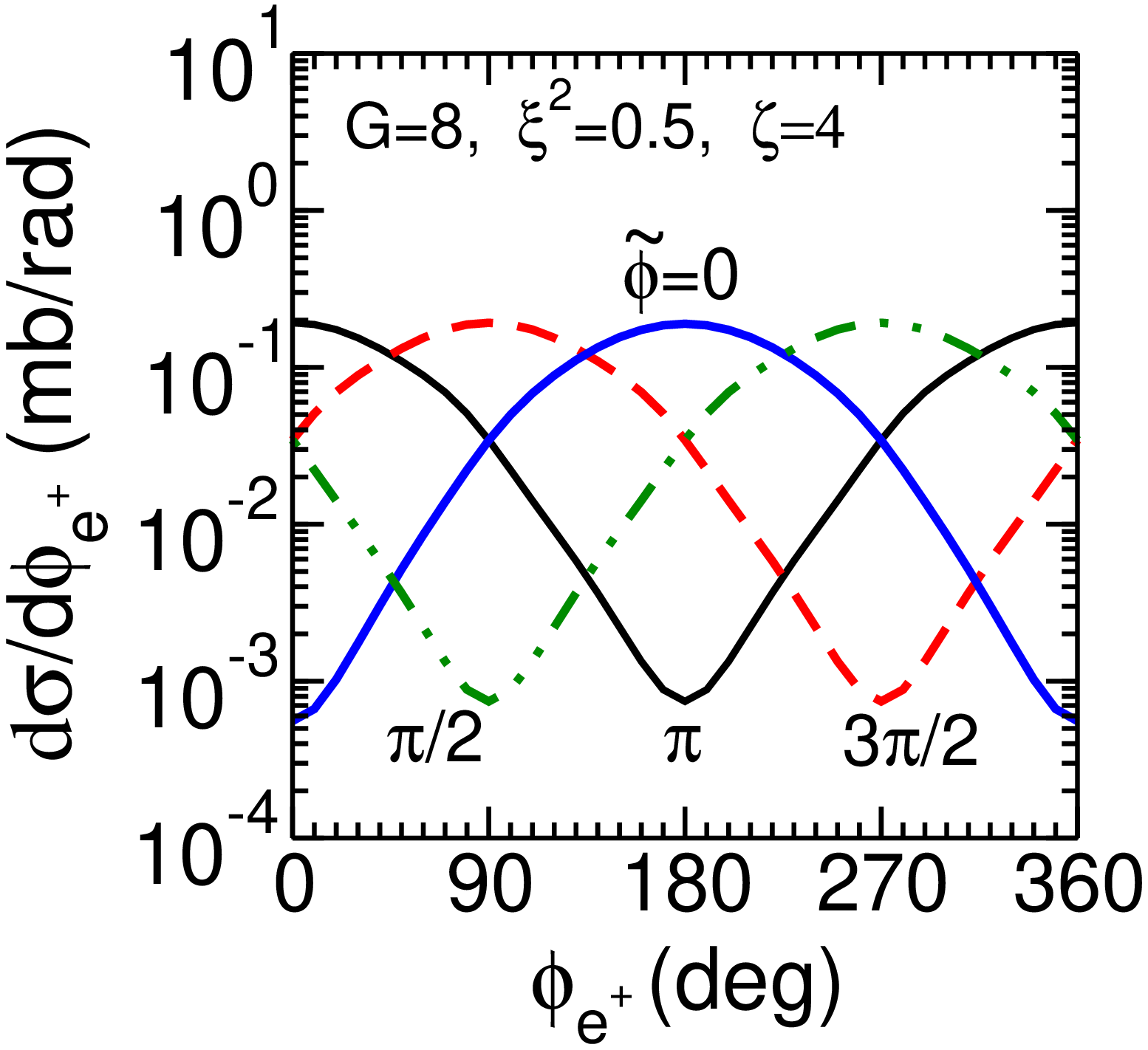}
\caption{\small (Color online)
Differential cross section $d\sigma/d\phi_{e^+}$ of Eq.~(\ref{III99})
for various values of the separation parameter $G$ and the carrier envelope phase $\tilde\phi$.
For $N = 1/2$, $\xi^2=0.5$ and $\zeta=4$.
Left panel: $G = 4$ (sold black curve), 5 (dashed green curve) and
10 (dashed red curve); $\tilde \phi = 0$.
Right panel: $G = 8$ and
$\tilde \phi = 0$ (solid blue curve), $\pi/2$ (dot-dashed green curve),
$\pi$ solid black curve), $3 \pi / 2$ ( dashed red curve).
When plotting the curves as a function of $\chi_{e^+}=\phi_{e^{+}}-\tilde\phi$
they coincide with the blue curve  labeled by $\tilde \phi = 0$.
\label{Fig:003}}
 \end{figure}

To be specific we consider the differential cross section
 $d\sigma/d\phi_{e^+}$ for sub-cycle pulse with $N = 1/2$
 as a function of azimuthal angle
 of positron momentum at different values of separation parameter
 $G$ and fixed CEP $\tilde\phi=0$ exhibited in the left panel of Fig.~\ref{Fig:003}.
One can see the oscillating structure of the cross
section. The positions of the maxima and the frequencies of
the oscillations depend on the separation parameter.

The dependence of the differential cross section for ultra-short pulse
on CEP at fixed separation parameter $G=8$ is expressed
in the right panel of Fig.~\ref{Fig:003}
One can see that the cross sections have a bump-like structure
and, besides, the bump positions coincide with the
corresponding carrier phases. If one plots the cross sections
as a function of the useful  "scale variable"
\begin{eqnarray}
\chi_{e^+}=\phi_{e^{+}}-\tilde\phi
\label{chi}
\end{eqnarray}
then the all curves expressed in right panel
of Fig.~\ref{Fig:003} merge into one
(that is the blue one labeled by $\tilde \phi = 0$).
The entire dependence the differential cross section on ˜$\tilde\phi$ is contained
in the variable $\chi_{e^+}$ (cf.~\cite{CEPTitov}). Later we use this variable
in our discussion.

According to (\ref{phi1}) the differential cross section would be enhanced at
 \begin{eqnarray}
 \chi_{e^+}=\phi^{\rm max}_{e^{+}}-\tilde\phi=(2n +1)\pi\pm G\Delta
 \label{phi1}
 \end{eqnarray}
 This is in pretty good agreement with results of our full
 calculation shown in the left panel of Fig.~\ref{Fig:003}
 (recollect that in this calculation $\tilde\phi=0$).
 Thus, for pulse separation $G=4$ choice for integer $n=0,\,-1$  leads to
 the bump position $\phi_{e^+}=\pi +(\pm\,2\pi)$ and $-\pi +(\pm\,2\pi)$.
 This is coincides with results exhibited in the
right panel of Fig.~\ref{Fig:003}, since
 the phase factor $(\pm \,2\pi)$ may be omitted. Similar results one can
 obtained for separation parameters
 $G=10$ with $\phi_{e^+}^{max}=0,\pm\,2\pi$ and $G=5$ with
 $\phi_{e^+}^{max}=\pm\,\pi/2\,\pm\,3\pi/2$, respectively.

\begin{figure}[htb]
 \includegraphics[width=0.48\columnwidth]{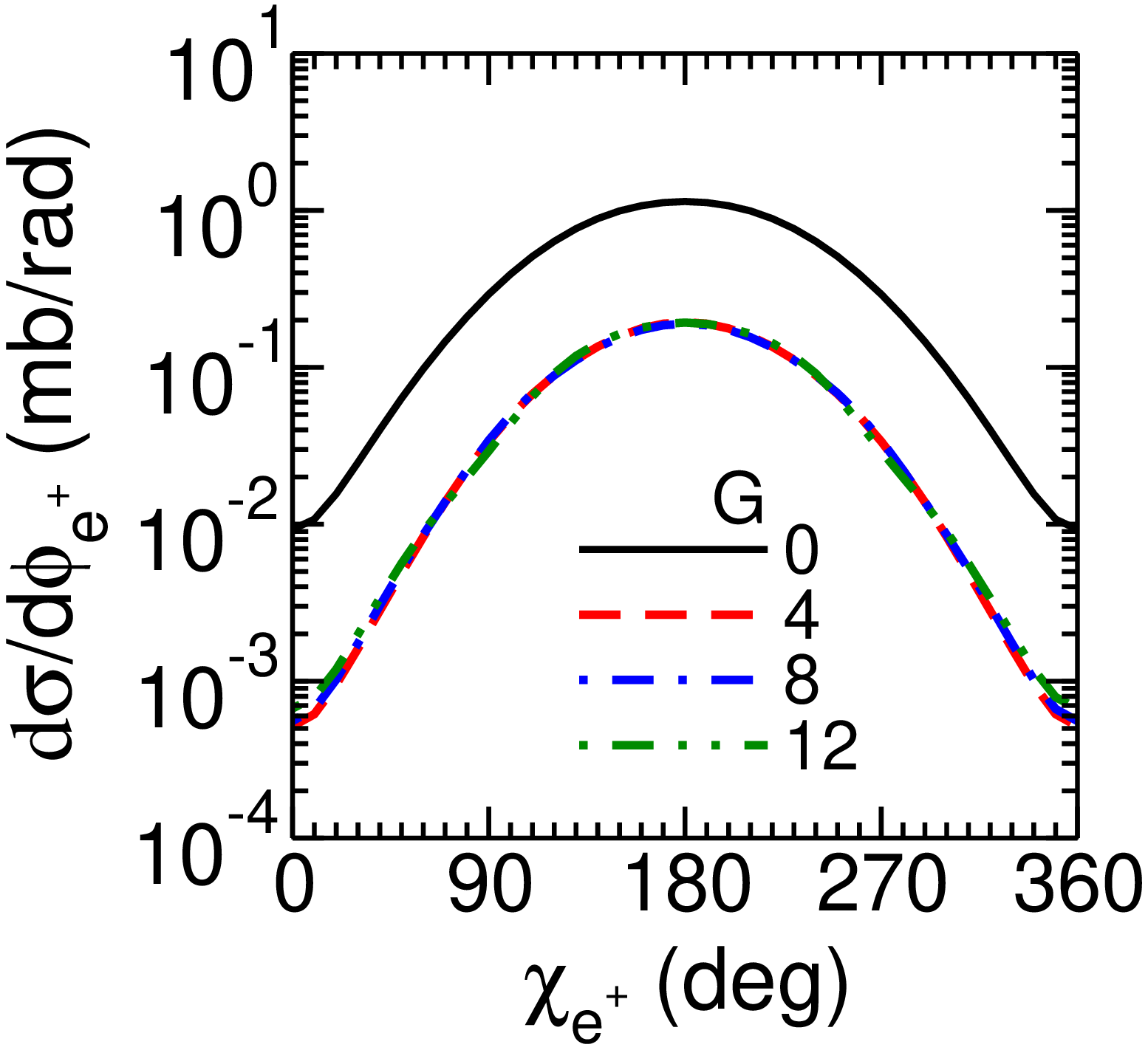} \hfill
 \includegraphics[width=0.48\columnwidth]{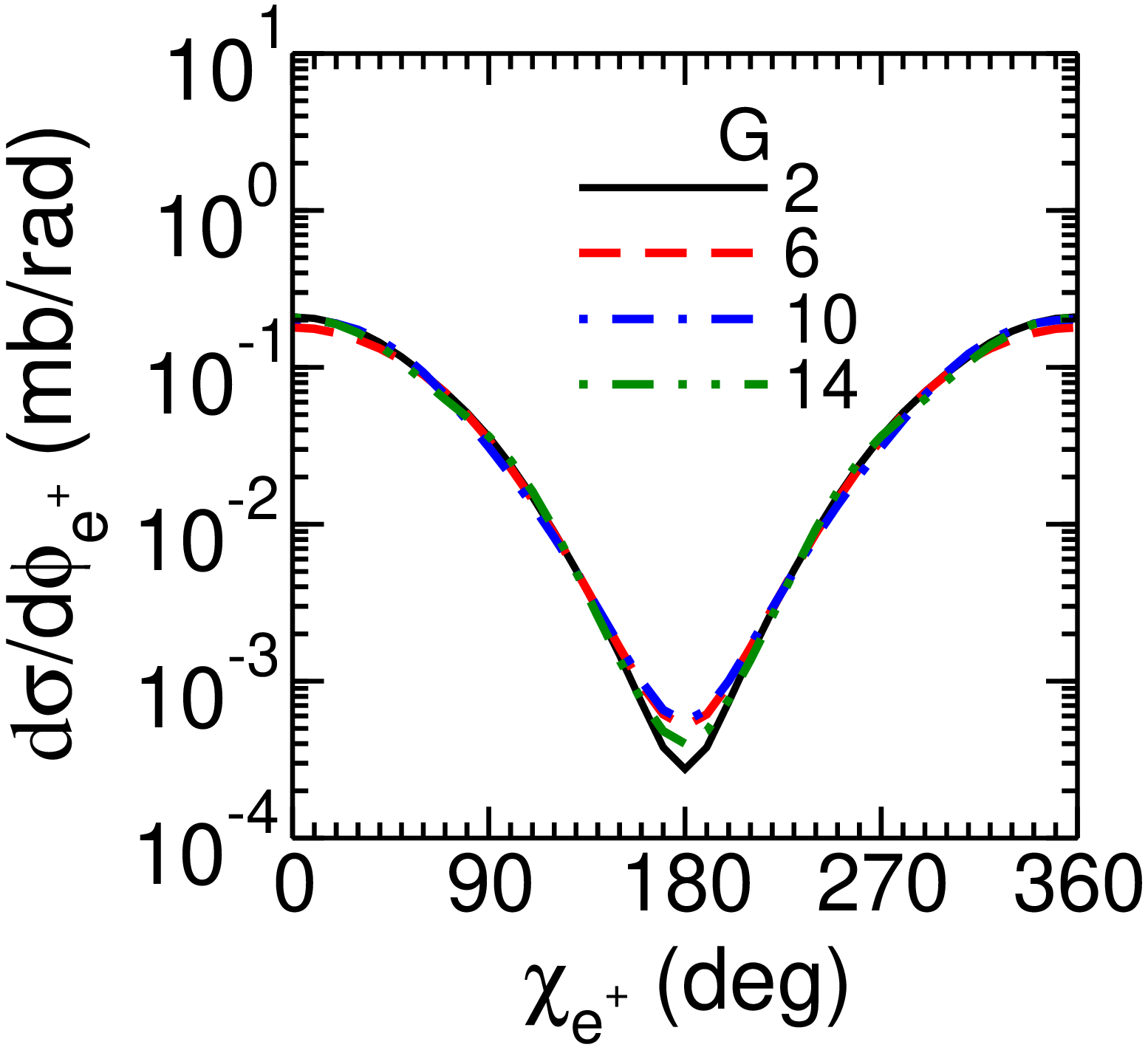} \\
 \includegraphics[width=0.48\columnwidth]{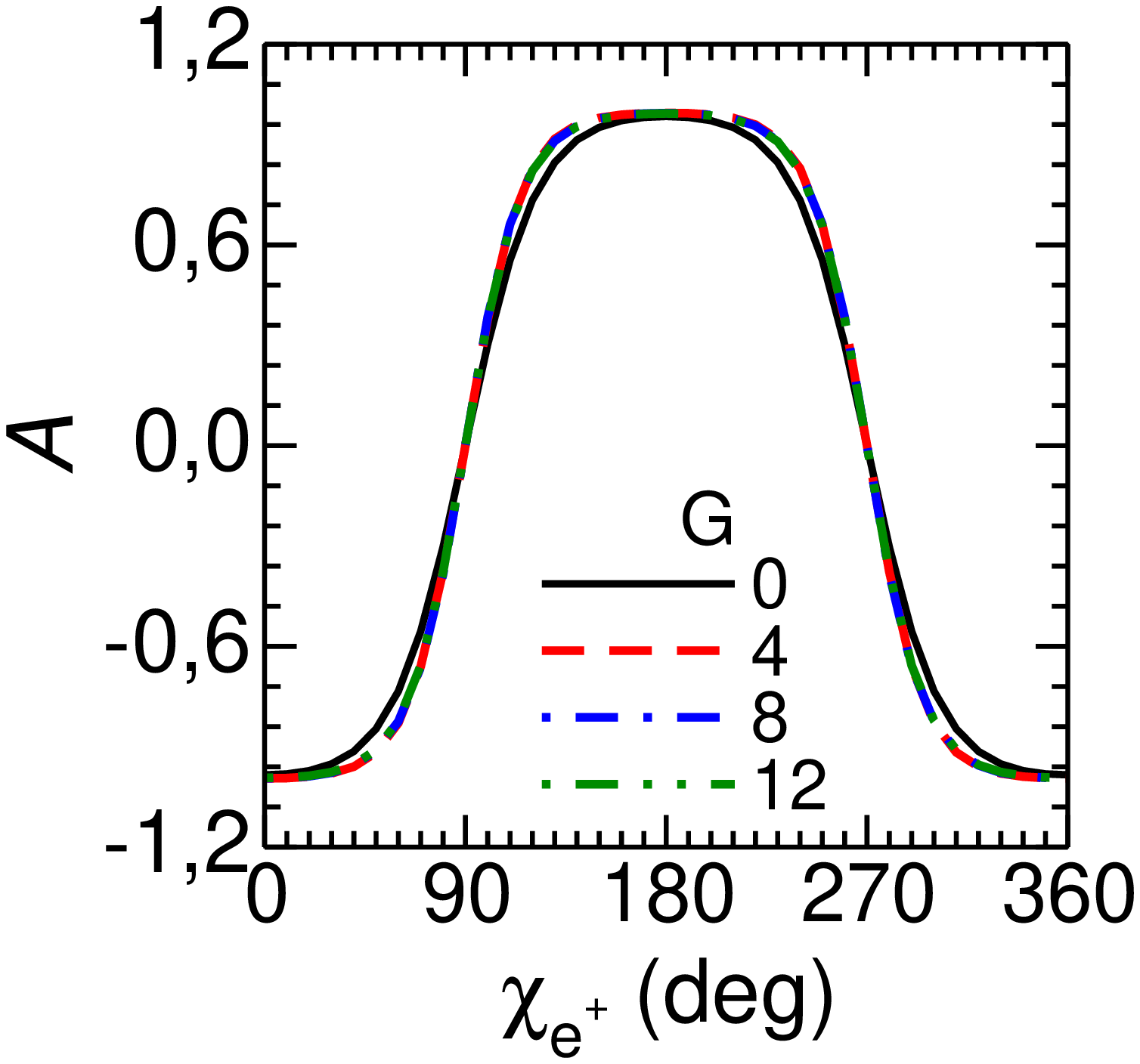} \hfill
 \includegraphics[width=0.48\columnwidth]{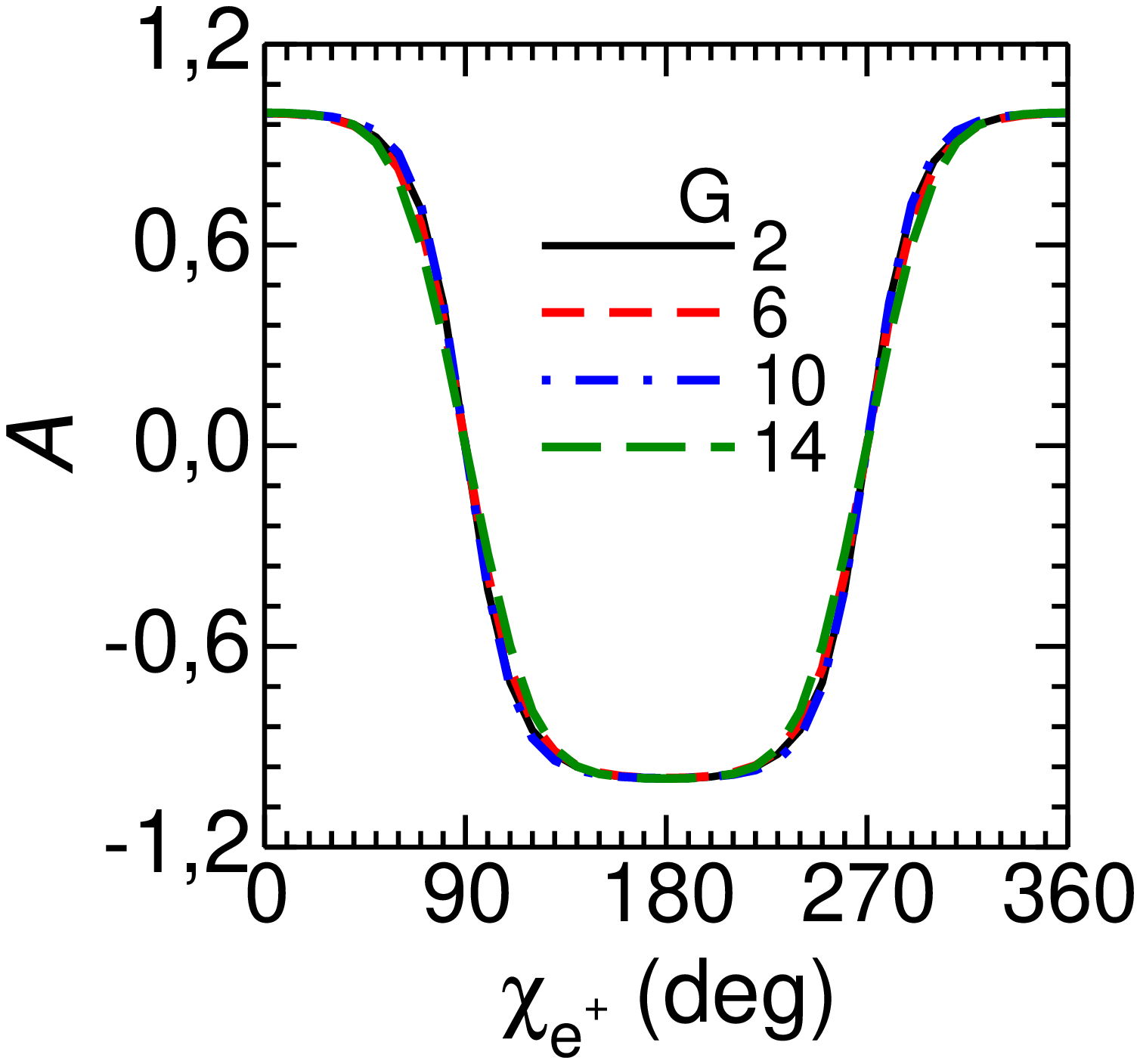}
 \caption{\small (Color online)
 Top panels :
  The differential cross section of Eq.~(\ref{III99})
  as a function of  $\chi_{e^{+}}$
 for $N = 1/2$, pulses separation parameter
 $G=0, 4, 8, 12$ and $G=2, 6, 10, 14$ shown in the left and right panels,
 respectively.
 Bottom panels : The same as in the top panels, but for the anisotropy of Eq.~(\ref{U9}).
 \label{Fig:03}}
 \end{figure}

Let us now focus on the dependence of the azimuthal
angular distribution on the separation parameter $G$, again for sub-cycle
pulse with $N = 1/2$. At $G = 4n$, the cross section
of $\ee$ production is expected to be enhanced at
\begin{eqnarray}
\chi_{e^+} = (2n+1)\pi~.
\label{phi2}
\end{eqnarray}
The corresponding cross sections and anisotropies as a
function of $\chi_{e^+}$ calculated numerically using Eqs.~(\ref{III99}) and
(\ref{U9}, respectively are displayed in the top and bottom
parts of the right panels in Fig.~\ref{Fig:03}. One can see sharp
bumps in the cross sections and anisotropies at $\chi_e^{+}=\pi$,
which coincide with our qualitative prediction (\ref{phi2}).

The case $G = 0$ is exceptional, since the two pulses
completely overlap and the cross section is greatly enhanced.
The main reason of such enhancement is related to the modification
of highly oscillating function $\Psi$ in Eq.~(\ref{UU8}).
Now, the leading term reads
\begin{eqnarray}
\Psi\simeq \ell\phi
 &-& 2z\Delta \cos(\Phi)\int\limits_{-\infty}^{\frac{\phi}{\Delta}} dt\, f_0(t)\cos(t\Delta)
 \nonumber\\
&-& 2z\Delta  \sin(\Phi)\int\limits_{-\infty}^{\frac{\phi}{\Delta}} dt\, f_0(t)\sin(t\Delta),
 \label{UU88}
 \end{eqnarray}
which leads to renormalization $z\to 2 z$ or $\xi^2\to 4\xi^2$.
The cross section increases with $\xi^2$ see, for example~\cite{TitovPEPAN})
and our next section therefore one can see strong
enhancement of the cross section at $G=0$.

The situation changes dramatically
for the separation parameter $G=2+4n$.
In this case, the cross sections and anisotropies are  enhanced at
$\phi_{e^+}-\tilde\phi=0$ and $2\pi$
as illustrated in the right panels of Fig.~\ref{Fig:03}.
Here, we face also some (approximate) symmetry: the right column
curves are generated by shifting
$\chi_{e^+} \to \chi_{e^+} \pm \pi$.

Both examples show the explicit correlation
between separation parameter $G$ and the carrier
phase $\tilde \phi$ in the different cross section $d\sigma/d\phi_{e^+}$,
especially for $G<2$. For larger values of  $G$,
the cross section becomes independent of the separation parameter $G$.

\begin{figure}[htb]
 \includegraphics[width=0.48\columnwidth]{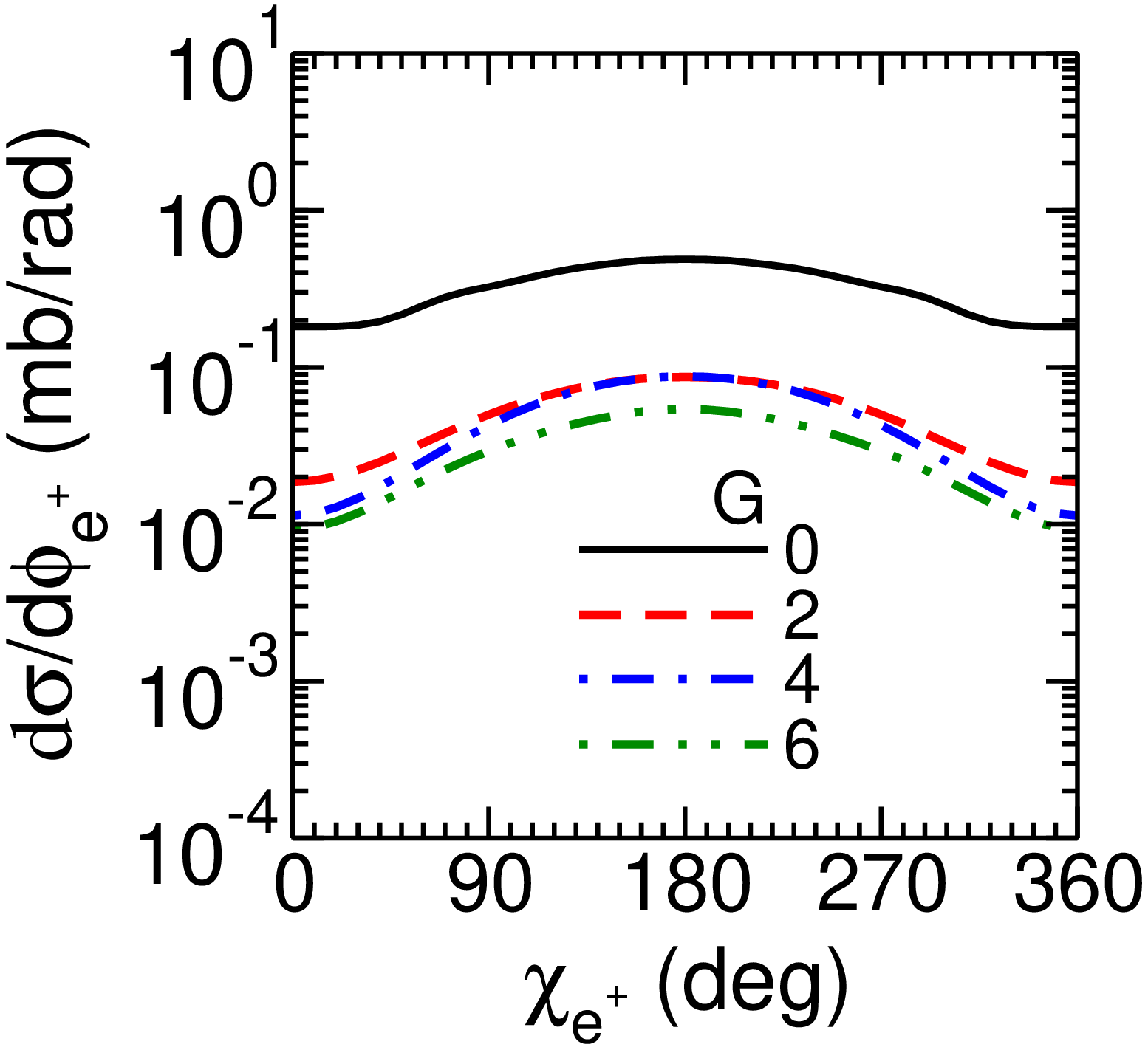} \hfill
 \includegraphics[width=0.48\columnwidth]{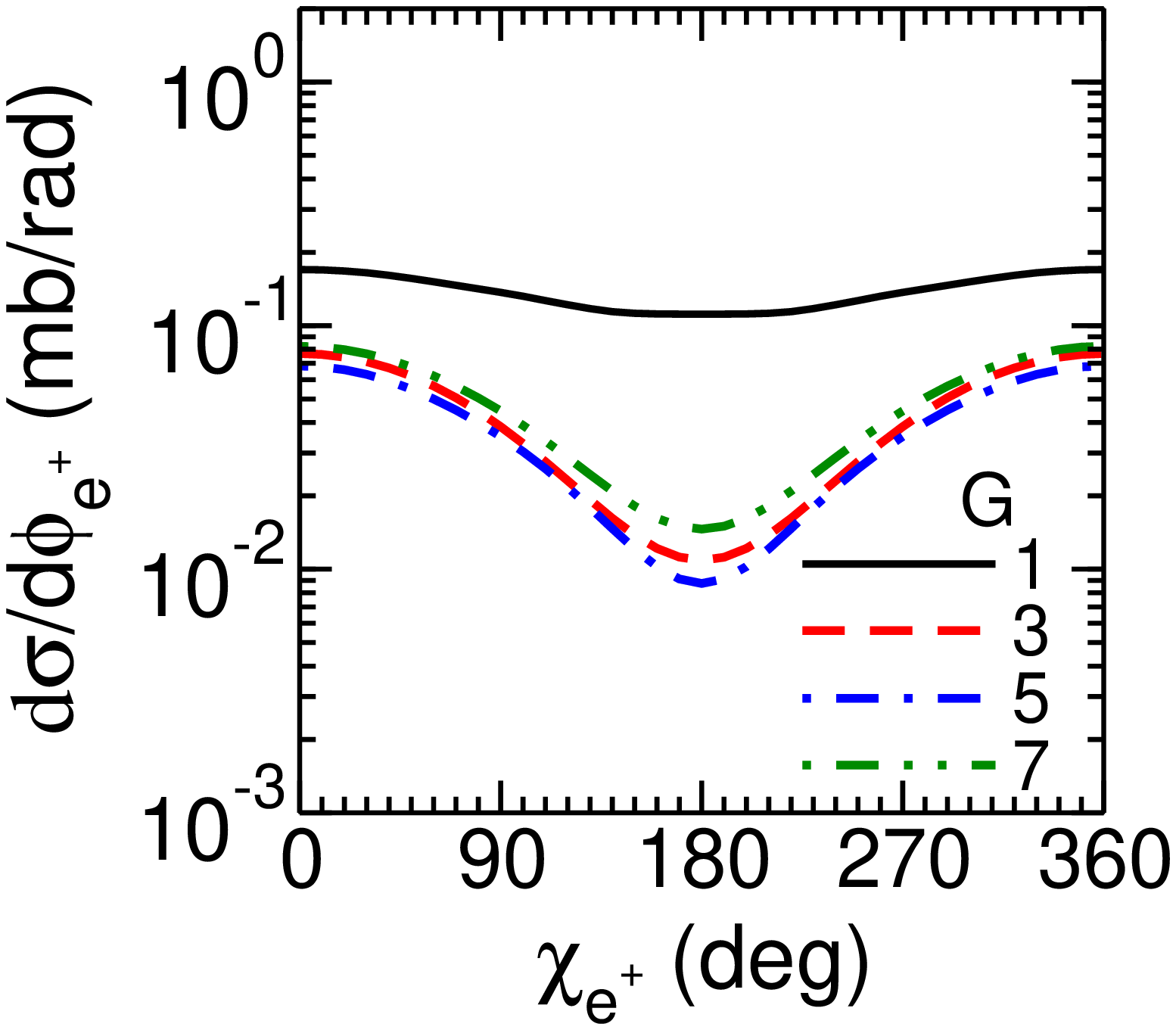}\\
 \includegraphics[width=0.48\columnwidth]{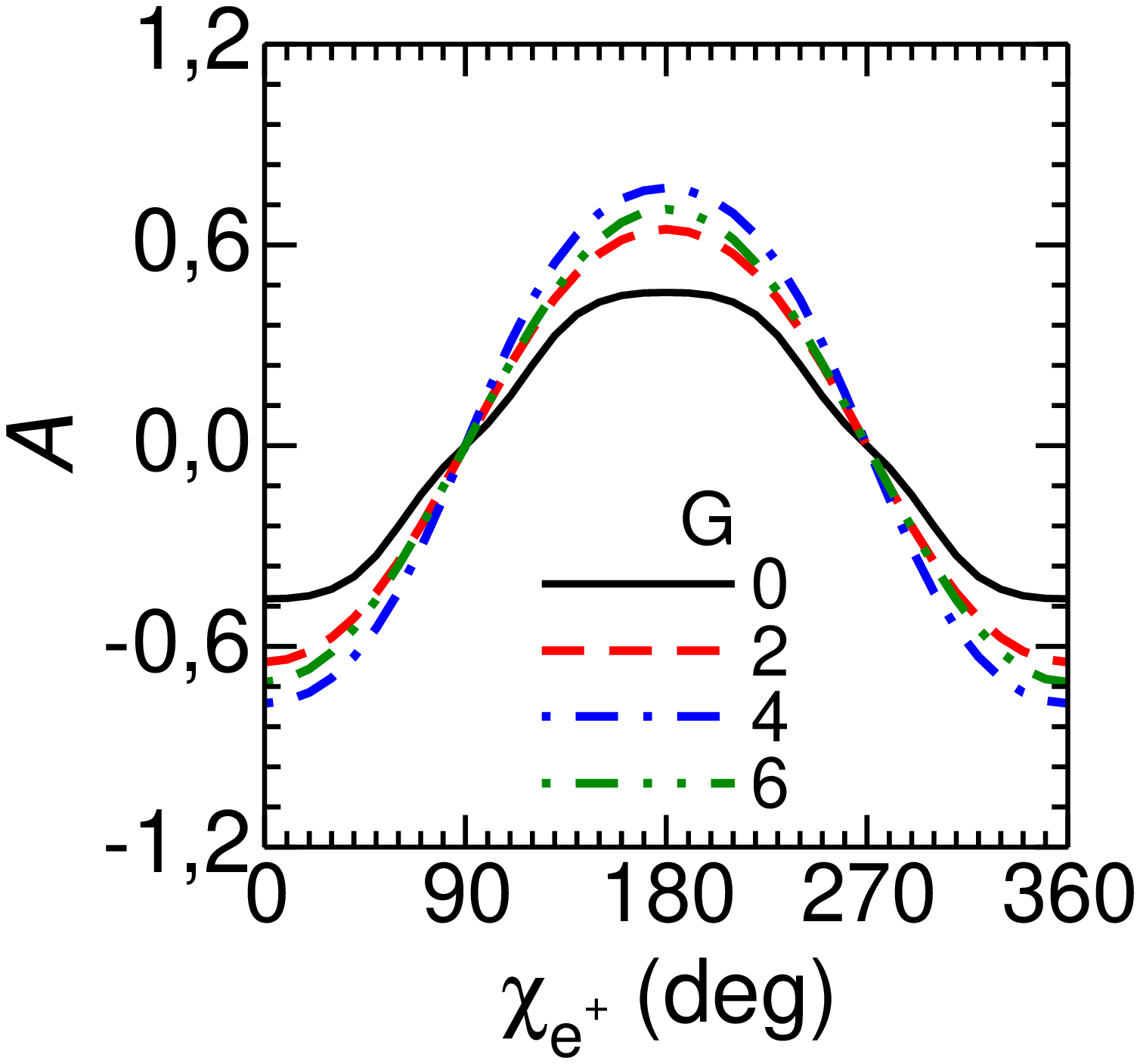} \hfill
 \includegraphics[width=0.48\columnwidth]{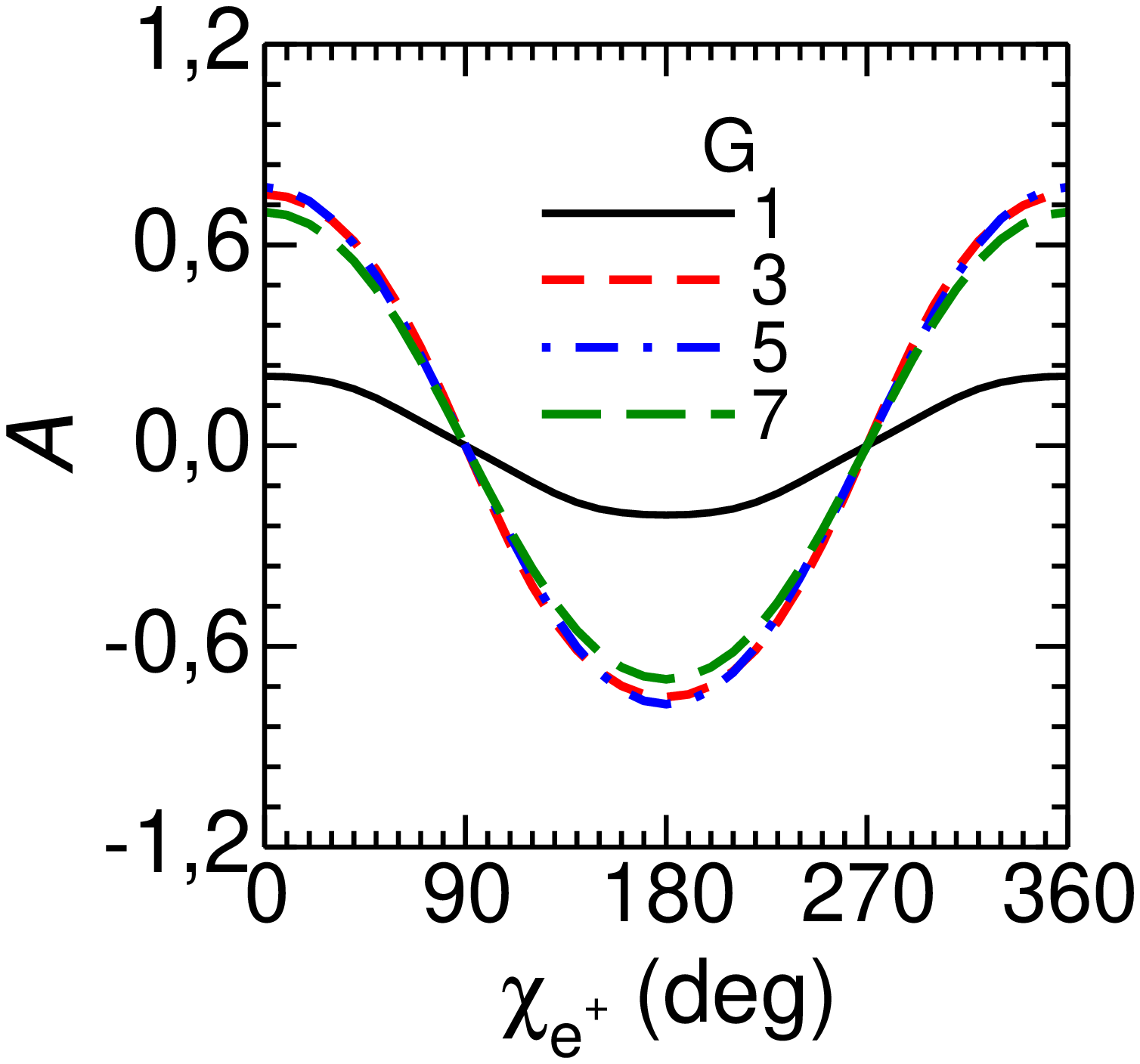}
 \caption{\small{(Color online)
 The same as in Fig.~\ref{Fig:03}
 but for $N=1$ and pulses separation parameter
 $G=0, 2, 4, 6$ and $G=1, 3, 5, 7$ shown in the left and right panels,
 respectively.
 \label{Fig:04}}}
 \end{figure}

A similar behavior is expected for wider pulses.
For instance, in Fig.~\ref{Fig:04} we display results for the short pulse
 with $N=1$ corresponding to $\Delta=\pi$.
 Using Eq.~(\ref{phi2}) one can find that the bump-like structure
 with the bump position $\chi_{e^+}=\pi$ appears
 at $G=2n$, as confirmed by our full numerical
 calculation exhibited in the left panels of Fig.~\ref{Fig:04}.
 The values of the separation parameter $G=2n+1$
 result in an enhancement of the cross section and anisotropies at
 $\chi_{e^+}=0$ and $2\pi$ (cf.\ right panels of Fig.~\ref{Fig:04}).

\begin{figure}[htb]
 \includegraphics[width=0.48\columnwidth]{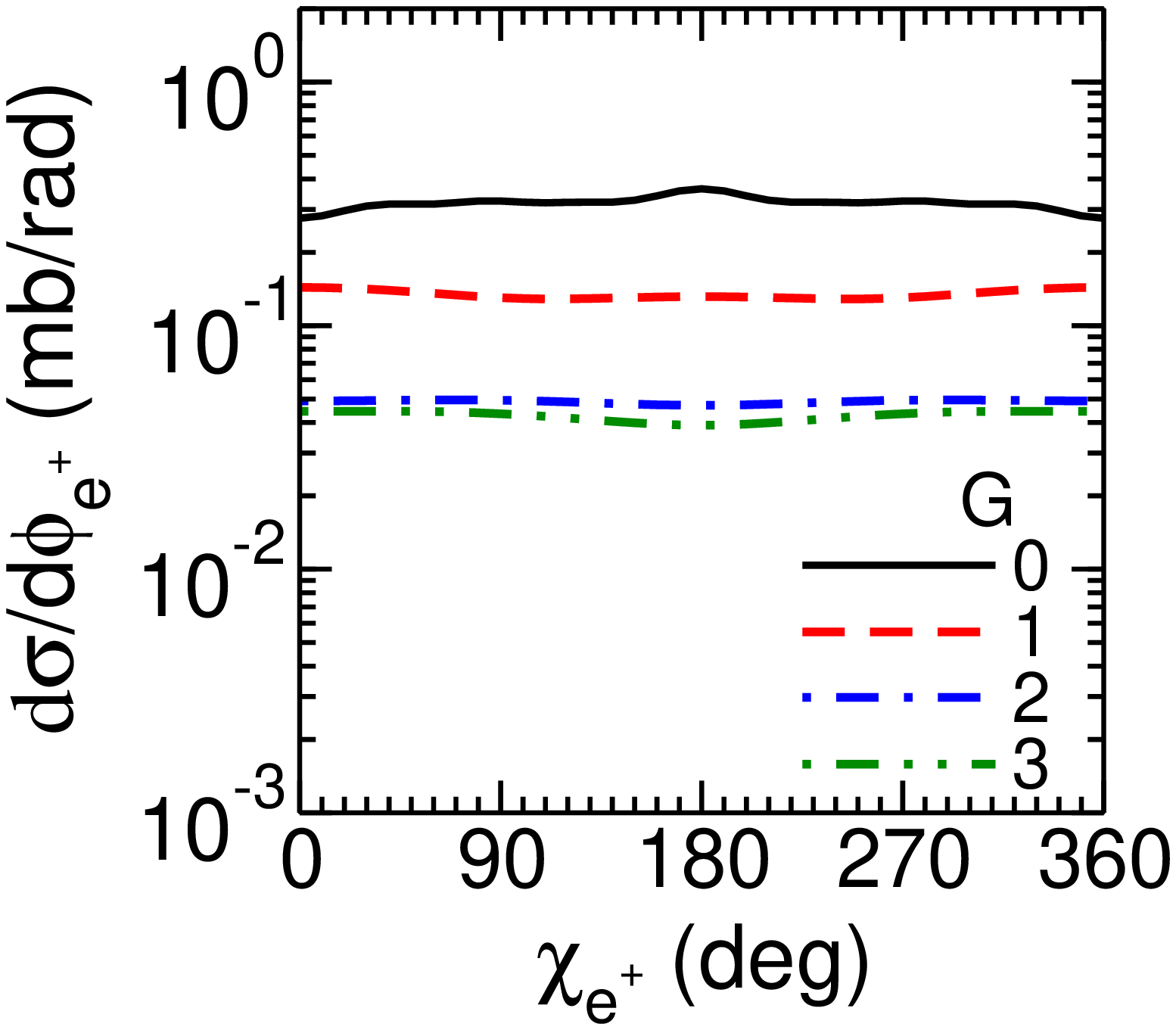}  \hfill
 \includegraphics[width=0.48\columnwidth]{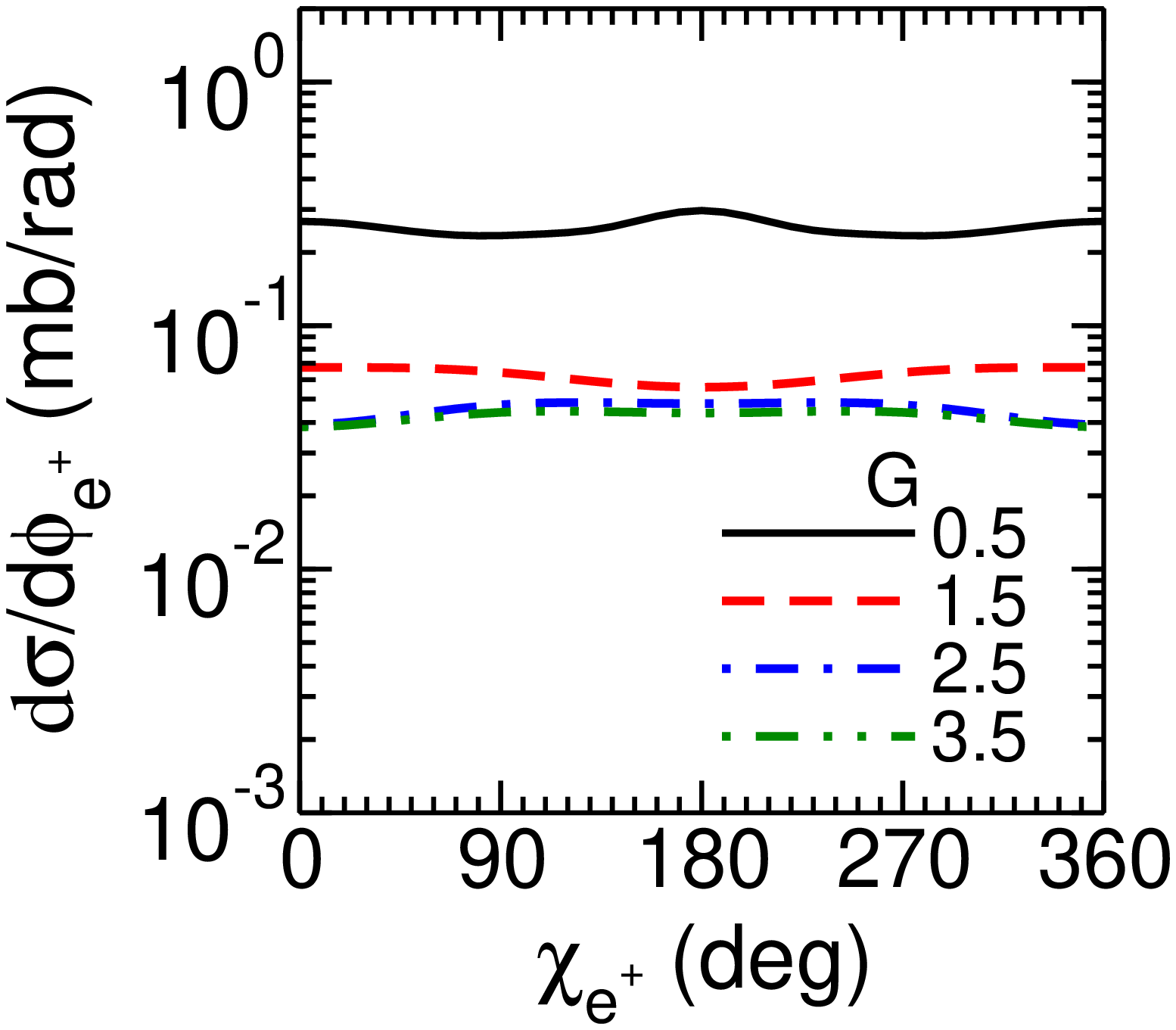}\\
 \includegraphics[width=0.48\columnwidth]{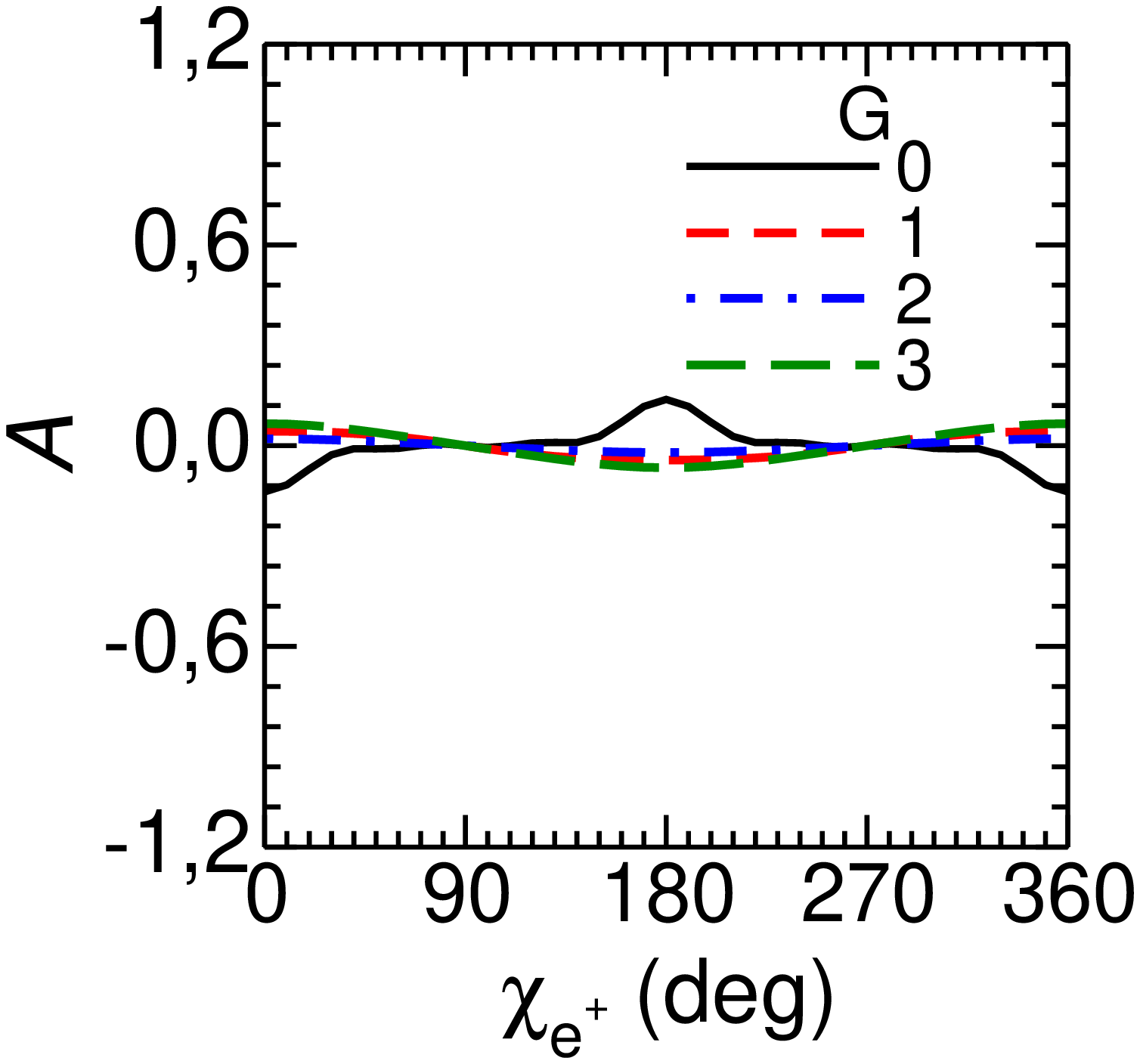}  \hfill
 \includegraphics[width=0.48\columnwidth]{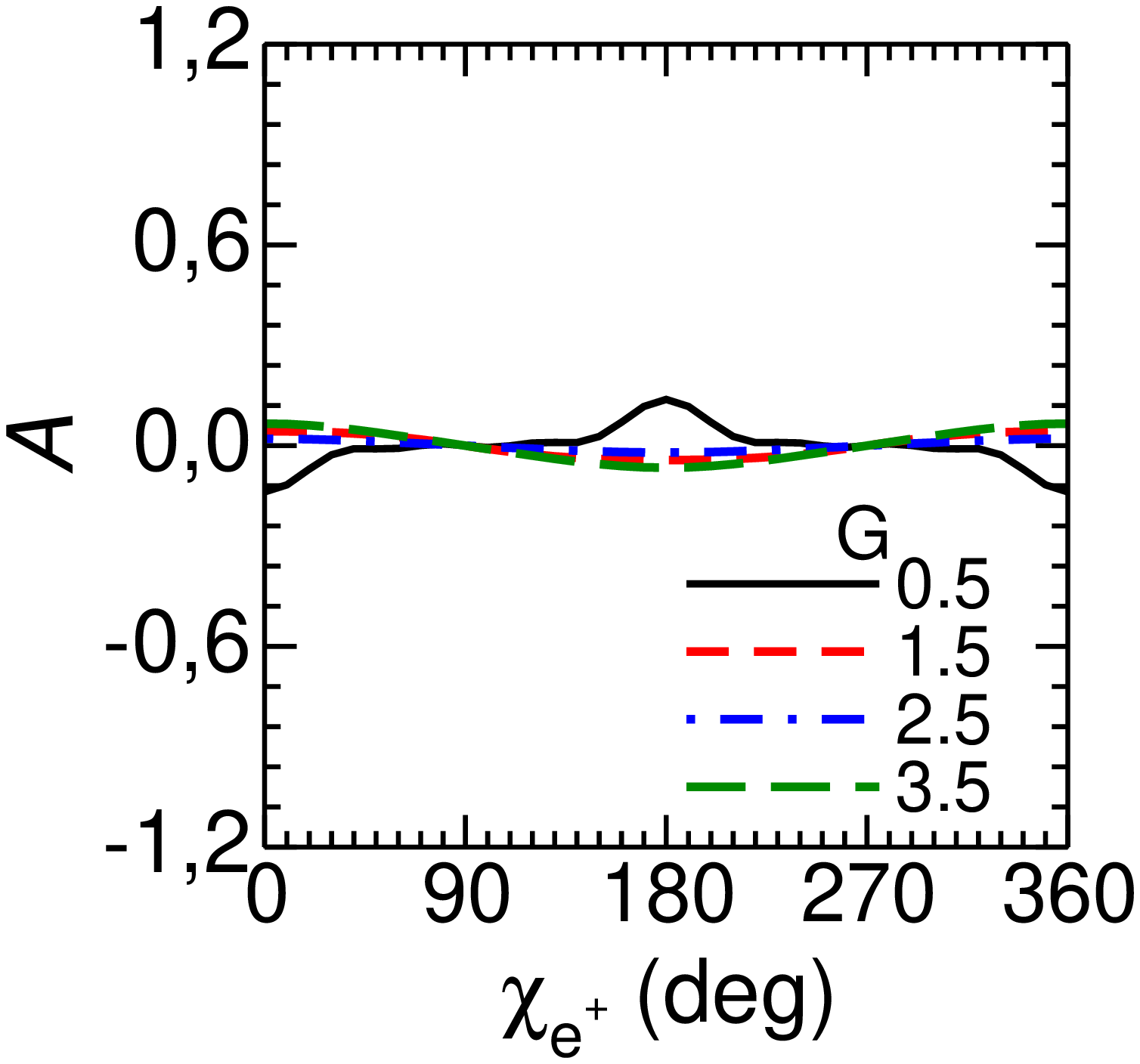}
 \caption{\small{
 (Color online)
 The same as in Fig.~\ref{Fig:03}
 but for $N=2$ and separation parameter
 $G=0, 1, 2, 3$ and $G=1/2, 3/2, 5/2, 7/2$ shown in the left and right panels,
 respectively.
 \label{Fig:05}}}
 \end{figure}

  For wider pulses with $N\geq 2$ the bump structure
  and the non-monotonic behavior becomes very weak.
  For example, Fig.~\ref{Fig:05} illustrates
  our result for the short pulses with $N=2$, that is $\Delta=2\pi$.

 Nevertheless, some enhancement
(weak bump structure at $\phi_{e^+}-\tilde\phi=\pi$) is expected at
$G=0 ,\,1, \,2, \,\cdots$, while the enhancement
at $\phi_e=0$ and $2\pi$ is expected at half-integer values
$G=1/2,\,3/2,\,5/2,\, \cdots$. Our corresponding result is depicted
in Fig.~\ref{Fig:05}. Our full numerical calculation does not support
the qualitative prediction of (\ref{UU8}).

\subsection{Total cross section}

With the given parameterization (1, 2), the total cross section depends on
the available energy $s$ (encoded in the sub-threshold parameter $\zeta$),
the laser intensity parameter $\xi$,
the double-pulse distance parameter $G$,
and the cycle number $N$ (encoded in the pulse width parameter $\Delta$):
$\sigma (s; \xi^2, G, N)$ since the CEP dependence disappears.
A few of these dependencies are now considered.
For simplicity and without loss of generality below we consider sub-cycle
and short pulses with $N=1/2$ and $N=1$, respectively.
Our analysis shows that wilder pulses do not bring
new qualitative results.

\begin{figure}[htb]
 \includegraphics[width=0.48\columnwidth]{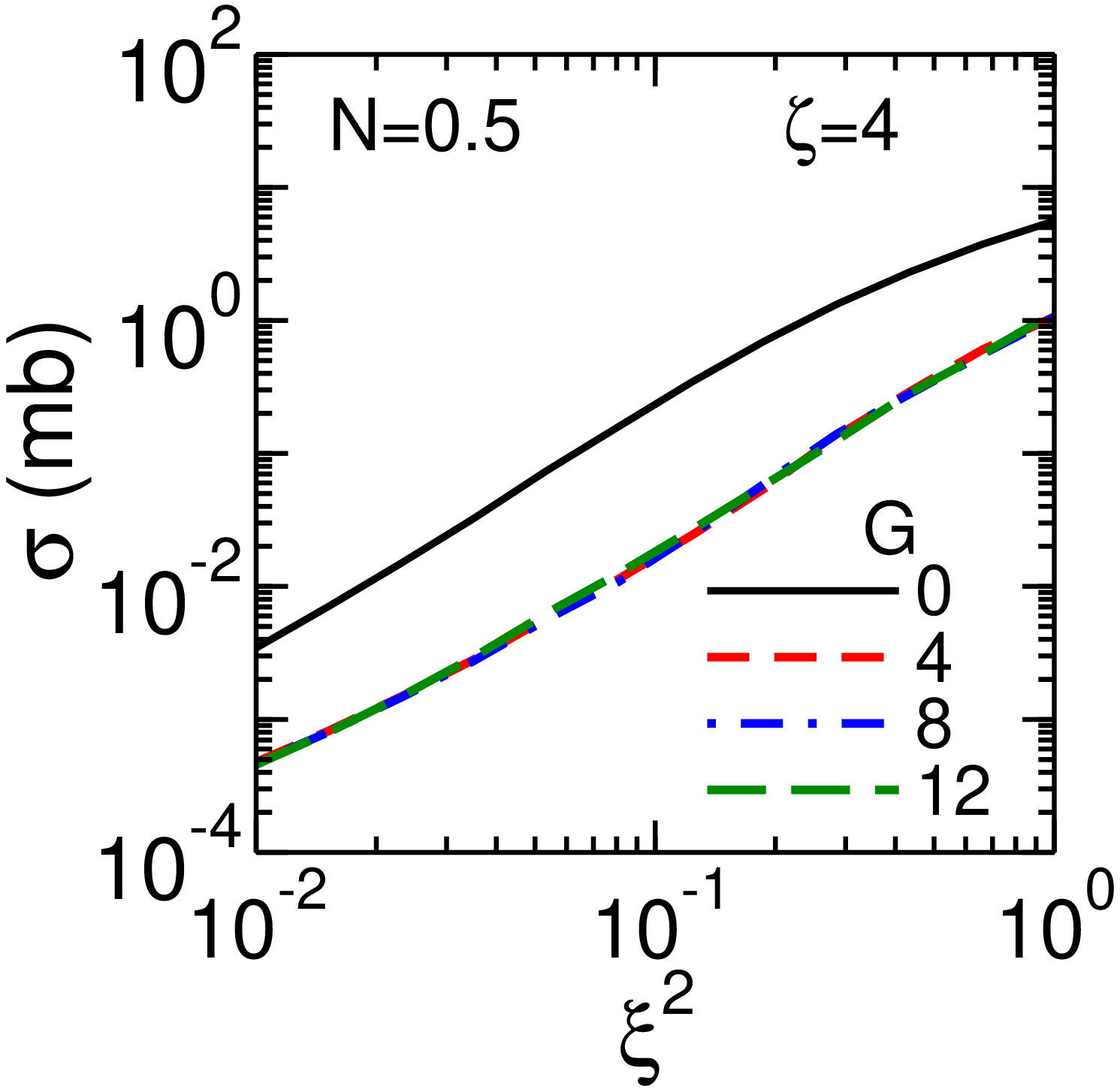} \hfill
 \includegraphics[width=0.48\columnwidth]{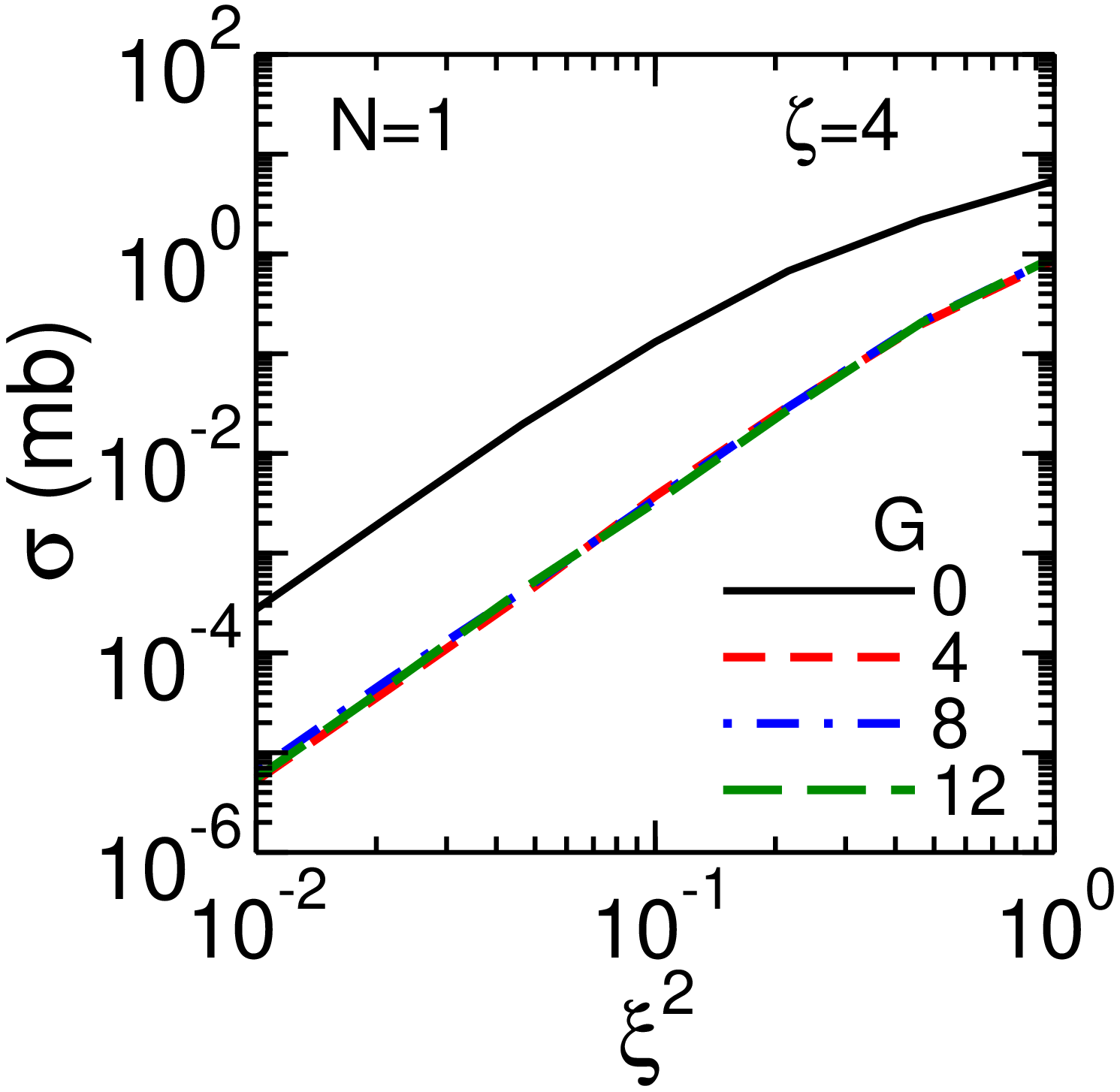}
 \caption{\small{
 (Color online)
  The total cross section of the $\ee$ production as a function
 of the reduced field intensity $\xi^2$. The results are for $N= 1/2$,
sub-threshold parameter $\zeta=4$
 and various values of the separation parameter $G$ listed in
 the legend.
 \label{Fig:06}}}
 \end{figure}

The dependence of the total cross section
%integrated over the azimuthal angle $\phi_{e^+}$
as a function of the field intensity $\xi^2$
is exhibited in Fig.~\ref{Fig:06}. We choose sub-threshold pulse
with $N=0.5$ %($\Delta=\pi/2$)
and the separation parameter $G=4n$
according to Fig.~\ref{Fig:03}, left panels.
One can see that all curves coincide with another at $G \geq 2$,
where the dependence on the separation parameter disappears.
Qualitatively, the behavior of the cross sections shown in the left
and right panels are similar, being enhanced for sub-cycle pulse,
especially for small vales of $\xi^2$.

\begin{figure}[htb]
  \includegraphics[width=0.48\columnwidth]{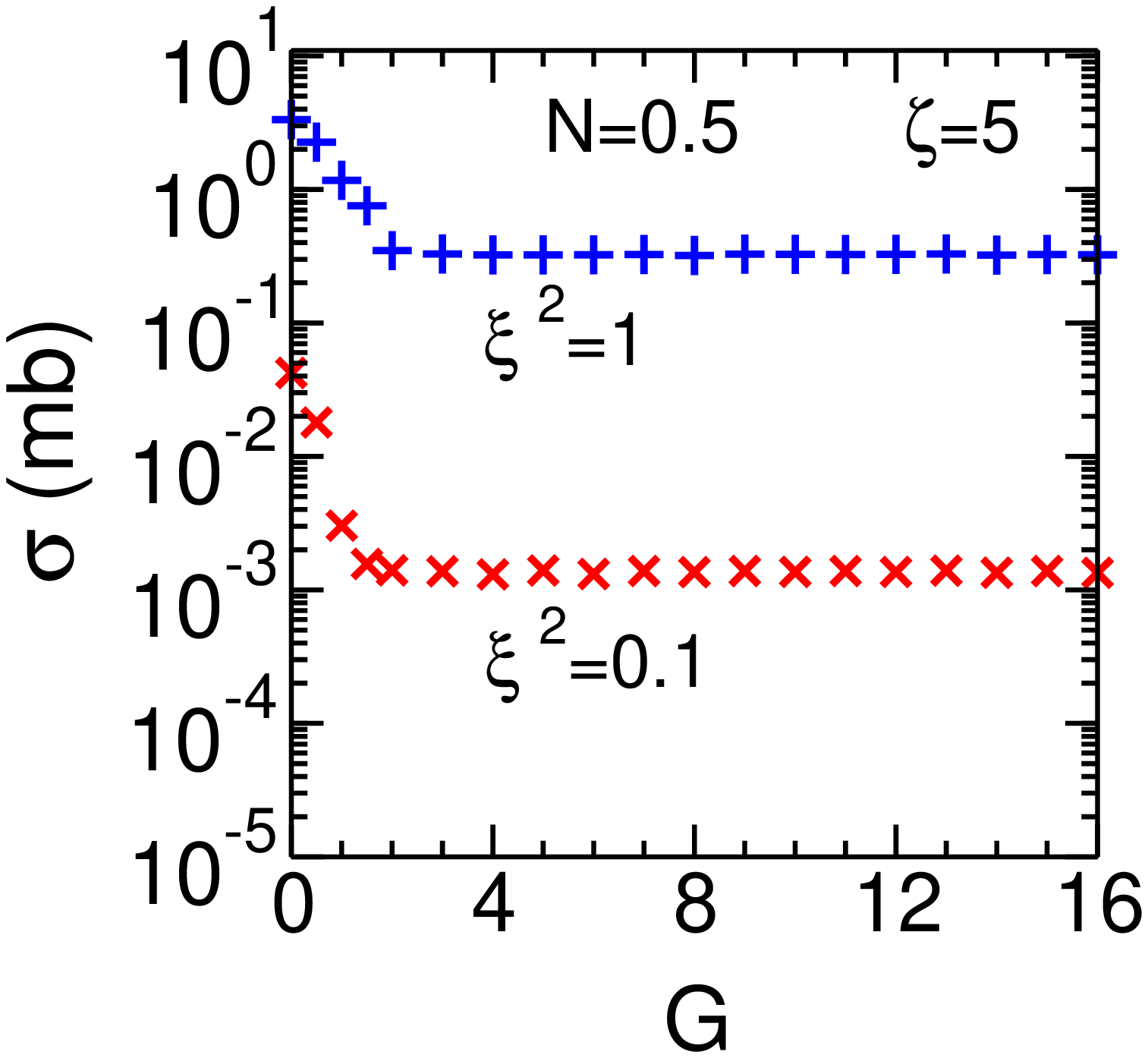}\hfill
  \includegraphics[width=0.48\columnwidth]{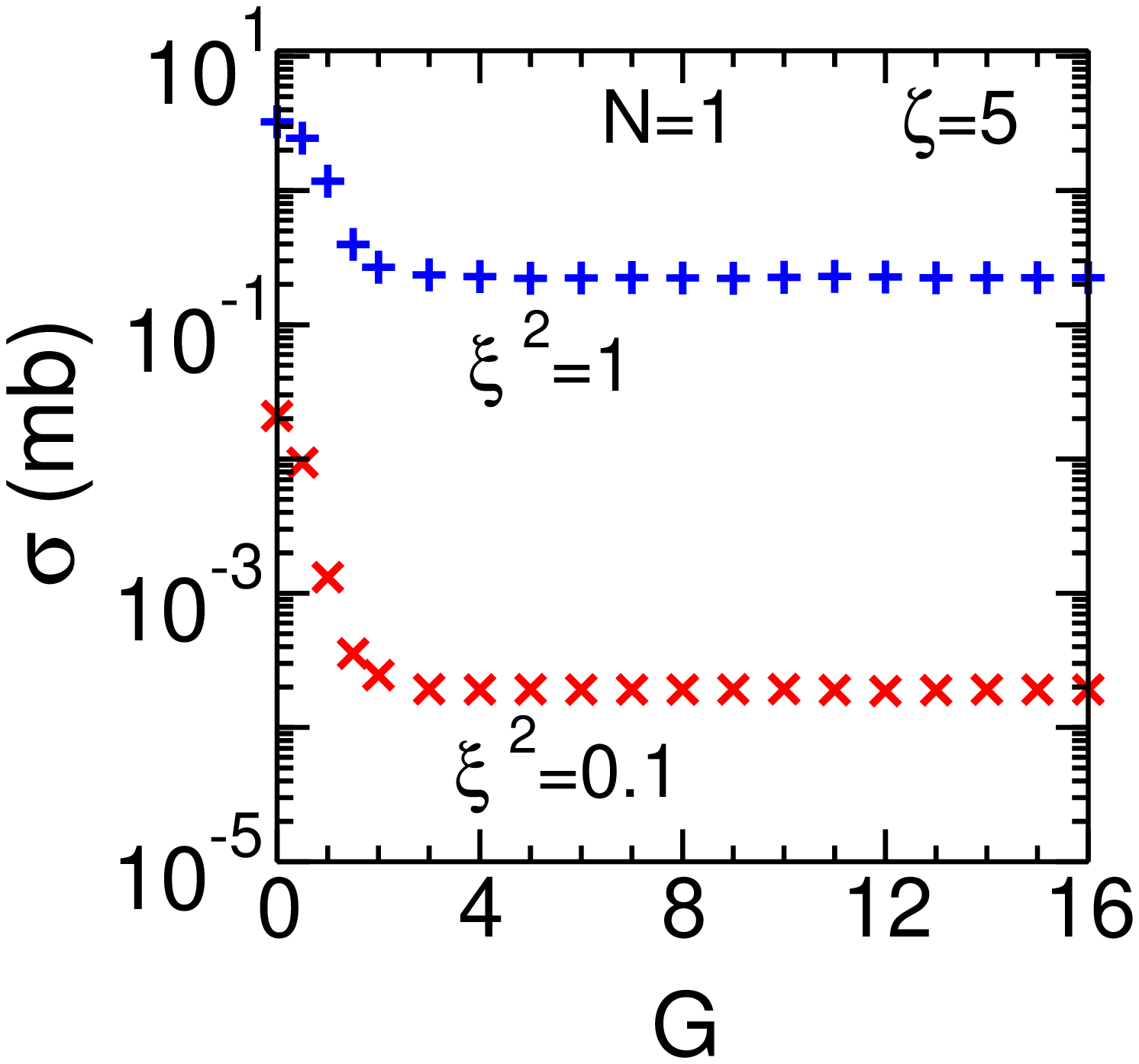}
 \caption{\small{
 (Color online)
  The total cross section of the $\ee$ production as a function
 of separation parameter $G$ at $\xi^2=0.1$ and $\xi^2=1$, and $\zeta=5$.
 Left and right panels correspond to
  $N = 1/2$ and $N=1$, respectively.
 \label{Fig:07}}}
 \end{figure}

This is correlated with results shown in Fig.~\ref{Fig:07},
where the total cross sections as a function of the separation
parameter $G$ for the sub-cycle ($N=1/2$) and short ($N=1$)
pules are shown in the left and right panels, respectively.
The calculations are for the
sub-threshold parameter $\zeta=5$ and the field
intensities $\xi^2=0.1$ and $1$. The cross sections for
$\xi^2=1$ are much greater compared to the case of
$\xi^2=0.1$.
One can see a significant enhancement of the
cross sections  at $G<2$ which is a consequence
of strong overlap of the two pulses, discussed above,
cf.\ top in
Figs.~\ref{Fig:03} and \ref{Fig:04}.
The cross sections at $G\geq 2$ are independent of $G$ and
exhibit some plateau. For the sub-cycle pulses
at chosen parameters within the accuracy of our calculation,
the cross sections at the plateau
are $\approx1.4 \times 10^{-3}$~mb and $\approx 0.32$~mb
for $\xi^2=0.1$ and $1$, respectively.
For the short pulses ($N=1$), the result is qualitatively
similar to that shown in the left panel
but the cross sections are smaller compared to the case of
$N=1/2$. Thus,  the cross sections at plateau are
$\approx 1.9 \times 10^{-4}$~mb and $\approx 0.22$ mb for $\xi^2=0.1$
and $1$, respectively. The
enhancement of the cross sections for the sub-cycle pulses
compared to the short pulses, especially for small $\xi^2$,
is consistent with our results exhibited
in Fig.~\ref{Fig:06} and below in Fig.~\ref{Fig:08}.
Note that the results displayed in
Figs.~\ref{Fig:06} and \ref{Fig:07} are obtained at different
values of the subthreshold parameter $\zeta=4$ and $5$, respectively.

\begin{figure}[htb]
 \includegraphics[width=0.48\columnwidth]{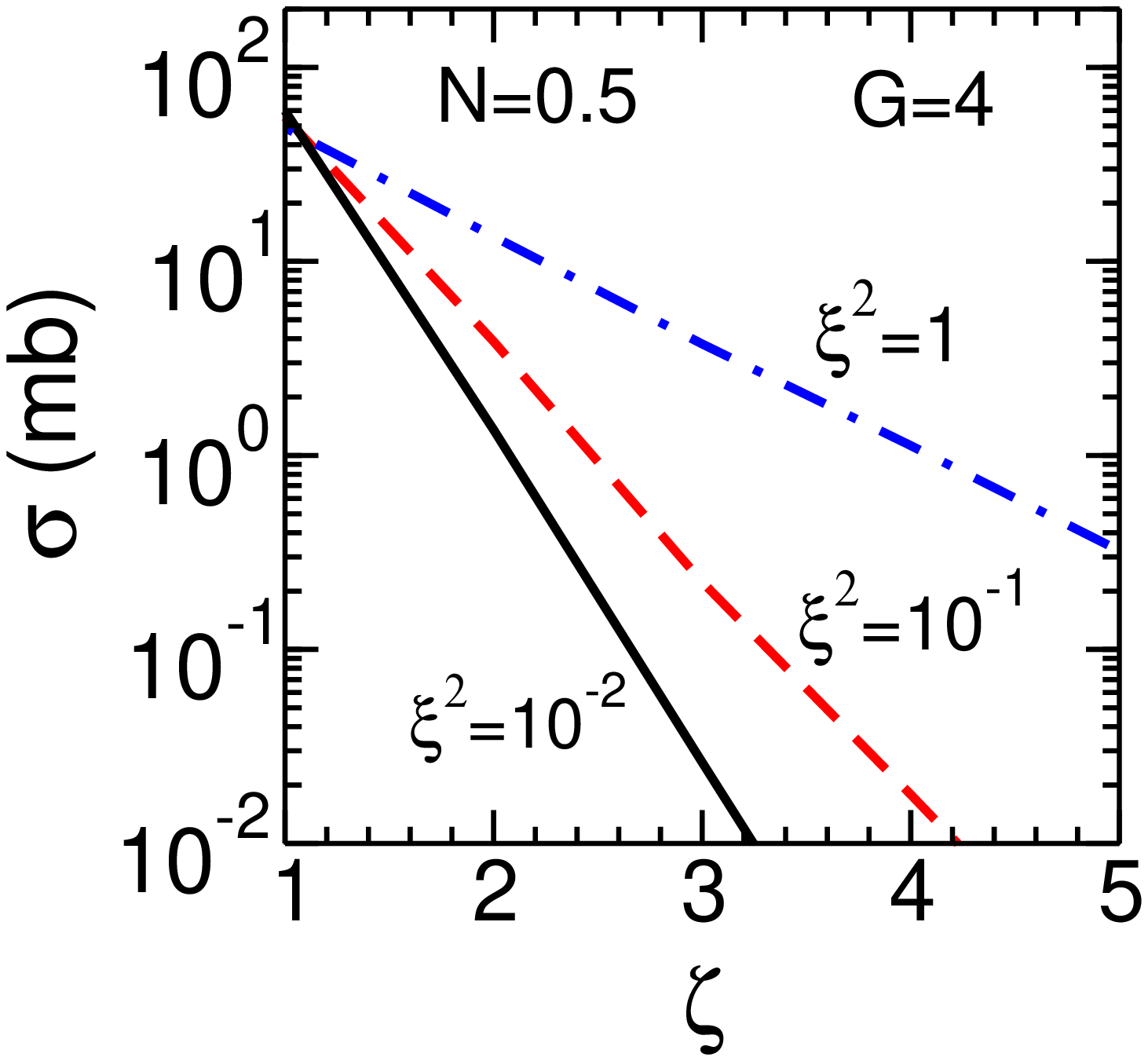} \hfill
 \includegraphics[width=0.48\columnwidth]{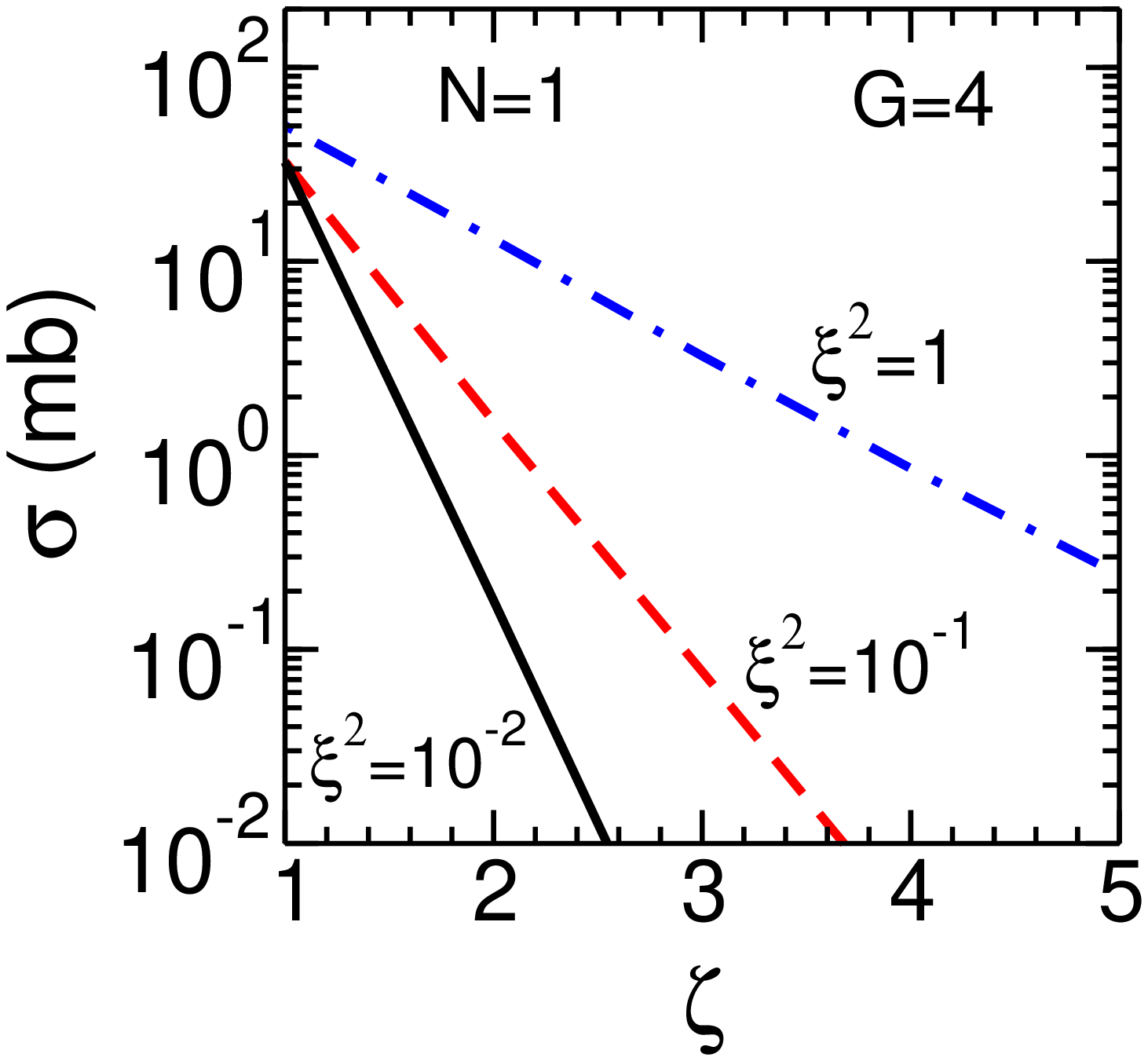}
  \caption{\small{
 (Color online)
  The total cross section of the $\ee$ production as a function
 of sub-threshold parameter $\zeta$ for various
 $\xi^2$ at fixed $G=4$ and $N = 1/2$.
 \label{Fig:08}}}
 \end{figure}

The total cross section as a function of the
sub-threshold parameter $\zeta$ is exhibited in
Fig.~\ref{Fig:08} for different values of field
intensity $\xi^2$ and $G=4$.
The left and right panels correspond to $N=1/2$ and $N=1$, respectively.
One can see an almost exponential decrease
of the cross section with $\zeta$.
The slope of the curves decreases with increasing
intensity of the electromagnetic field (or $\xi^2$).
(Closer inspection shows that this result does not depend
on  the separation parameter $G\geq 2$ which is correlated
with results exhibited in Fig.~\ref{Fig:07}.)
Qualitatively, results for sub-cycle and short pulses
are similar. The curves in case of sub-cycle
pulses are slightly flatter.

\section{summary}

In summary we have analyzed the non-linear (strong-field) Breit-Wheeler
$\ee$ production in case of
two consecutive circularly polarized short pulses with equal carrier phases.
We find that for short pulses
with the number of oscillation in the pulse $N\leq 1$
the azimuthal angular differential cross section %$d\sigma/d\phi_{e^+}$
is sensitive to both the carrier phase $\tilde\phi$ and
the separation parameter $G$. That  is
manifest in the bump-like behavior of the differential cross section.
Results of the full numerical calculations coincide with
simple analytical expressions.
This means that the azimuthal angle distribution of outgoing
fermions may be used as a powerful method for determination of the structure
of the short double pulses (or the carrier phase at the fixed pulse geometry).

The total cross section obeys a monotonic (near-exponential) increase with
the field intensity $\xi^2$ at fixed sub-threshold parameter $\zeta$
and an exponential decrease of cross section with increasing
$\zeta$ at fixed $\xi^2$.
The total cross sections increases significantly for  $G<2$
because of strong interference in case of overlapping pulses.
All these  facts can be used in planning appropriate experiments.

\begin{acknowledgments}
The authors gratefully acknowledge the collaboration with
D. Seipt, T. Nousch, T. Heinzl,
and useful discussions with
A. Ilderton, K. Krajewska,  M. Marklund, C. M\"uller, and R. Sch\"utzhold.
We thank S. Glenzer for pointing out the content of Ref.~\cite{Decker}
prior to publication.
The work is supported by R.~Sauerbrey and T.~E.~Cowan w.r.t.\ the study
of fundamental QED processes for HIBEF.
\end{acknowledgments}

\appendix

\section{Evaluation of the partial probability $\mathbf{w(\ell)}$}

Here, for simplicity we choose the carrier phase $\tilde\phi$ equal to zero.
Generalization for a finite $\tilde\phi$ is done in a straight forward manner.
The starting point is (cf.~\cite{TitovPRL,TitovPEPAN,CEPTitov})
\begin{eqnarray}
 \frac{d\sigma}{d\phi_{e^+}}=\frac{\alpha^2\,\zeta }{8m^4\xi^2N_0}\,
 \int\limits_{\zeta}^\infty \,d\ell \, v \int\limits_{-1}^{1} d\cos\theta_{e^+}\,
 |M_{fi}(\ell,u)|^2~, %\quad(\rm E1)
 \label{EQ1}
 \end{eqnarray}
where
\begin{eqnarray}
M_{fi}(\ell)=\sum\limits_{i=0}^3  [\bar u_{p'}\,\hat M^{(i)}\,v_p ]\,C^{(i)}(\ell)~,
\label{EQ2}
\end{eqnarray}
with
\begin{eqnarray}
 \hat M^{(0)}&=&\fs\varepsilon'~,\quad
 \hat M^{(1)}=-
  \frac{ e^2a^2 \,
 (\varepsilon'\cdot k)\,\fs k}
 {2(k\cdot p)(k\cdot p')}~,\nonumber\\
 \hat M^{(2,3)}&=&\frac{e\fs a_{(1,2)}\fs k\fs
 \varepsilon'}{2(k\cdot p')} - \frac{e\fs \varepsilon'\fs k\fs
 a_{(1,2)}}{2(k\cdot p)}~,
\label{Diracstructures}
\end{eqnarray}
where $u_{p'}$ and $v_{p}$ are the Dirac spinors of the electron and positron,
respectively, and
$\varepsilon'$ is the polarization four-vector of the probe photon $X$.

The functions $C^{(i)}(\ell)$ read
\begin{eqnarray}
C^{(0)}(\ell)&=&\frac{1}{2\pi}\int\limits_{-\infty}^{\infty}
d\phi \,{\rm e}^{i\ell\phi -i{\cal P(\phi)}}~,\nonumber\\
C^{(1)}(\ell)&=&\frac{1}{2\pi}\int\limits_{-\infty}^{\infty}
d\phi f^2(\phi)\,{\rm e}^{i\ell\phi -i{\cal P(\phi)}}~,\nonumber\\
C^{(2)}(\ell)&=&\frac{1}{2\pi}\int\limits_{-\infty}^{\infty}
d\phi f(\phi)\,\cos\phi\,{\rm e}^{i\ell\phi -i{\cal P(\phi)}}~,\nonumber\\
C^{(3)}(\ell)&=&\frac{1}{2\pi}\int\limits_{-\infty}^{\infty}
d\phi f(\phi)\,\sin\phi\,{\rm e}^{i\ell\phi -i{\cal P(\phi)}}~,
\label{A4}
\end{eqnarray}
with the phase function ${\cal P(\phi)}$ from (\ref{CP2}).
The integrand of the function $C^{(0)}$ does not contain the envelope
function $f(\phi)$ and therefore
it is divergent. One can "regularize" it by using the
prescription of  \cite{BocaFlorescu} yielding
\begin{eqnarray}
C^{(0)}(\ell) &=&\frac{1}{2\pi \ell}
\int\limits_{-\infty}^{\infty} d\phi
\left[ z\cos(\phi-\phi_0)\,f(\phi)-\xi^2\zeta u\,f^2(\phi)\right] \nonumber \\
&\times& {\rm e}^{i\ell\phi -i{\cal P(\phi)}}
+ \delta(\ell)\,{\rm e}^{-i{\cal P}(0)}~.
\label{A7}
\end{eqnarray}
The divergence is isolated in the last term. However, it does not
contribute because of kinematic considerations implying $\ell > 0$.

Utilizing Eqs.~(\ref{Diracstructures}), (\ref{A4}) and (\ref{A7})
and using the notation $Z_{ij}=\frac14{\rm Tr}[ M^{(i)}\,{M^{(j)}}^\dagger]$
one can express the mod-square of (\ref{EQ2}) in the following form
\begin{eqnarray}
\label{A8}
&&\frac14|M_{fi}(\ell)|^2 = \\
&&|C^{(0)}|^2Z_{00} + |C^{(1)}|^2Z_{11} + |C^{(2)}|^2Z_{22} + |C^{(3)}|^2Z_{33} \nonumber\\
&& + 2{\rm Re}\,C^{(0)}{C^{(1)}}^*Z_{01}+2{\rm Re}\,C^{(0)}{C^{(2)}}^*Z_{02} \nonumber \\
&& \hspace*{3.1cm} +2{\rm Re}\,C^{(0)}{C^{(3)}}^*Z_{03} \nonumber \\
&& + 2{\rm Re}\,C^{(1)}{C^{(2)}}^*Z_{12}+2{\rm Re}\,C^{(1)}{C^{(3)}}^*Z_{13}
\nonumber\\
&&+ 2{\rm Re}\,C^{(2)}{C^{(3)}}^*Z_{23}~. \nonumber
\end{eqnarray}
The averaging and sum over the spin variables in the initial and the final states
is already executed.
Trace calculations lead to
\begin{eqnarray}
Z_{00}&=&2\left(p\cdot p'+2m^2\right),
\nonumber\\
Z_{22}&=&Z_{33}=\frac{\xi^2m^2\left((k\cdot p)^2 + (k\cdot p')^2\right)}
{(k\cdot p) (k\cdot p')},
\nonumber\\
Z_{01}&=&\xi^2m^2,
\nonumber\\
Z_{02}&=&\frac{e\left(k\cdot p + k\cdot p'\right)
\left((p'\cdot a_1)(k\cdot p) - (p\cdot a_1)(k\cdot p')\right)}
{(k\cdot p) (k\cdot p')},
\nonumber\\
Z_{03}&=&Z_{02}\left(a_1\to a_2 \right),
\nonumber\\
Z_{11}&=& Z_{12}=Z_{13}=Z_{23}=0
\label{A9}
\end{eqnarray}
which can be expressed through more convenient variables by
\begin{eqnarray}
Z_{00}&=&2m^2\left(2u_\ell +1\right),
\nonumber\\
Z_{22}&=&Z_{33}=2\xi^2m^2(2u-1),
\nonumber\\
Z_{01}&=&\xi^2m^2,
\nonumber\\
Z_{02}&=&-2m^2u_\ell \frac{z}{\ell}\cos\phi_0,
\nonumber\\
Z_{03}&=&-2m^2u_\ell \frac{z}{\ell}\sin\phi_0
\label{A10}
\end{eqnarray}
%\underline{with $u_0=\ell/\zeta$.}
making Eq.~(\ref{A8}) surprizingly simple:
\begin{eqnarray}
\label{A12}
\frac14|M_{fi}(\ell)|^2&=&
2m^2(2u_0+1)|C^{(0)}|^2  \\
&+& 2m^2\xi^2(2u-1)\left(|C^{(2)}|^2 + |C^{(3)}|^2\right)
\nonumber\\
&+& 2 \xi^2 m^2 C^{(0)}{C^{(1)}}^* \nonumber \\
&-& 4 m^2u_0 \frac{z}{\ell}
\left(C^{(2)}\cos\phi_0 + C^{(3)}\sin\phi_0 \right){C^{(0)}}^*.
\nonumber
\end{eqnarray}

In principle, using the explicit expressions Eqs.~(\ref{A4}), (\ref{CP2}) and (\ref{A7}))
for the functions $C^{(i)}$ one can evaluate
$\frac14|M_{fi}(\ell)|^2$ numerically.
However, it turns out that it is more convenient
to use another representation for the square of the matrix element,
which allows to carry out a qualitative analysis
of the partial probability. For this aim we introduce new functions
 $Y_\ell$ (\ref{Yl}) and $X_\ell$ (\ref{Xl})  as well as  $\tilde Y_\ell$ (\ref{CP3})
which may be considered as
the generalized Bessel functions for the finite e.m. pulse
which allow to express the functions $C^{(i)}$ as
 \begin{eqnarray}
\label{A14}
C^{(0)}(\ell)&=&\widetilde Y_\ell(z){\rm e}^{i\ell\phi_0}~, \\
%\widetilde Y_\ell(z)&=&\frac{z}{2\ell} \left(Y_{\ell+1}(z) + Y_{\ell-1}(z)\right) - \xi^2\frac{u}{u_\ell}\,X_\ell(z)~,\nonumber\\
C^{(1)}(\ell)&=&X_\ell(z)\,{\rm e}^{i\ell\phi_0}~,\nonumber\\
C^{(2)}({\ell})&=&\frac{1}{2}\left( Y_{\ell+1}{\rm e}^{i(\ell+1)\phi_0}
+ Y_{\ell-1}{\rm e}^{i(\ell-1)\phi_0}\right) ~,\nonumber\\
C^{(3)}({\ell})&=&\frac{1}{2i}\left( Y_{\ell+1}{\rm e}^{i(\ell+1)\phi_0}
- Y_{\ell-1}{\rm e}^{i(\ell-1)\phi_0}\right)~. \nonumber
\end{eqnarray}

The identity
\begin{eqnarray}
 C^{(2)}(\ell)\cos\phi_0 +C^{(3)}(\ell)\sin\phi_0 &=&
 \frac12 \left( Y_{\ell+1}  + Y_{\ell -1} \right) \nonumber\\
&\times& {\rm e}^{i\ell\phi_0}
\label{A15}
\end{eqnarray}
and Eqs.~(\ref{A14}), together with some
cumbersome algebra, lead to the final result (\ref{III26-0}) due to
$w(\ell) \equiv \frac{1}{4m^2}|M_{fi}(\ell)|^2$.
%\begin{eqnarray}
%w(\ell) \equiv \frac{1}{4m^2}|M_{fi}(\ell)|^2
%2 |\widetilde Y_\ell(z)|^2+\xi^2(2u-1)\nonumber\\
%&\times&\left(|Y_{\ell-1}(z)|^2 + |Y_{\ell+1}(z)|^2 -2
%{\rm Re}\,(\widetilde Y_\ell(z)X^*_\ell(z))\right)~.
%\label{A16}
%\end{eqnarray

\end{document}